\documentclass[letterpaper,useAMS,usenatbib]{mn2e}
\usepackage[colorlinks=true,
            linkcolor=blue,
            urlcolor=blue,
					  citecolor=blue]{hyperref}
\usepackage{amssymb}
\usepackage{graphicx}
\usepackage{amsmath}
\usepackage[amssymb]{SIunits} 
\usepackage{booktabs}
\usepackage{hhline}
\usepackage{breqn}
\usepackage{standalone}
\usepackage{dcolumn}
	\newcolumntype{d}[1]{D{.}{.}{#1}}
\usepackage{tabularx}
\usepackage{booktabs}
\usepackage{microtype}
\usepackage{placeins}
\graphicspath{{graphics/}}
\defcitealias{D13}{D13}
\defcitealias{Jee13}{J13}
\defcitealias{M12}{M12}
\defcitealias{Sifon13}{Sif\'{o}n 2013}

\providecommand{\adsurl}[1]{\href{#1}{ADS}}
 
\title[The return of the merging galaxy subclusters of El Gordo?]
{The return of the merging galaxy subclusters of El Gordo?}
\author[Karen Y. Ng et al.]{Karen Y. Ng,$^{1}$
	William A. Dawson,$^{2}$ 
	D. Wittman,$^{1}$
	M. James Jee,$^{1}$ 
	John P. Hughes,$^{3}$ 
	\newauthor
	Felipe Menanteau,$^{4, 5}$
	Crist\'{o}bal Sif\'{o}n$^{6}$\\
$^{1}$Department of Physics, University of California Davis, One Shields
Avenue, Davis, CA 95616, USA\\ 
$^{2}$Lawrence Livermore National Laboratory, P.O. Box 808, Livermore, CA 94551-0808, USA \\
$^3$Department of Physics \& Astronomy,
Rutgers University, 136 Frelinghysen Rd., Piscataway, NJ 08854, USA\\
$^{4}$National Center for Supercomputing Applications, University of
Illinois at Urbana-Champaign, 1205 W. Clark St, Urbana IL, 61801, USA\\
$^{5}$Department of Astronomy, University of Illinois at Urbana-Champaign,
W. Green Street, Urbana, IL 61801, USA\\
$^{6}$Leiden Observatory, Leiden University, PO Box 9513, NL-2300 RA
Leiden, Netherlands\\}
\begin{document}
\date{arXiv 1412.1826} \pagerange{\pageref{firstpage}--\pageref{lastpage}}
\pubyear{2014} \maketitle \label{firstpage}
\begin{abstract} 
Merging galaxy clusters with radio relics provide rare insights to the merger
dynamics as the relics are created by the violent merger process. 
We demonstrate one of the first uses of the
properties of the radio relic to reduce the uncertainties of the dynamical variables 
and determine the 3D configuration of a cluster merger, ACT-CL J0102-4915, nicknamed El Gordo. 
From the double radio relic observation and the X-ray observation of a
comet-like gas morphology induced by motion of the cool core, 
it is widely believed that El Gordo is observed shortly after the first
core-passage of the subclusters.
We employ a Monte Carlo simulation to investigate the three-dimensional (3D)
configuration and dynamics of El Gordo. 
Using the polarization
fraction of the radio relic, we constrain the estimate of the
angle between the plane of the sky and the merger axis to be $\alpha = 21\degree
\pm^9_{11}$. We find the relative 3D merger speed of El Gordo to be
$2400\pm^{400}_{200}~\kilo\meter~\second^{-1}$ at pericenter. The two possible
estimates of the time-since-pericenter are $0.46\pm^{0.09}_{0.16}$ Gyr and
$0.91\pm^{0.22}_{0.39}$ Gyr for the outgoing and returning scenario respectively.
We put our estimates of
the time-since-pericenter into context by showing that if the time-averaged
shock velocity is approximately equal to or smaller than the pericenter velocity of the
corresponding subcluster in the 
center of mass frame, the two subclusters are more likely to be moving towards, rather
than away, from each other, post apocenter. 
We compare and contrast the merger scenario of El Gordo with that of the Bullet
Cluster, and show that this late-stage merging scenario 
explains why the southeast dark matter lensing peak of El Gordo is
closer to the merger center than the southeast cool core. 
%Finally, we provide our insight on what information from simulations and
%observations would help us better constrain the merger scenarios for other
%bimodal merging clusters. 
\end{abstract}
\section{Introduction} 
Mergers of dark-matter-dominated galaxy clusters probe properties
of the cluster components like no other systems. 
In terms of mass content, dark matter makes up $\sim80\%$ of the mass of clusters
of galaxies, a smaller portion of the mass consist of intercluster
gas($\sim15\%$ in mass content), and sparsely spaced galaxies ($\sim2\%$ in mass content). During a merger of
clusters of the high mass range of $10^{14} M_{\sun}$ to the low mass end of
$10^{15} M_{\sun}$, the subclusters are accelerated to high speeds of 
$\sim 2000 - 3000~\kilo \meter~\second^{-1}$ (\citealt{Lage2014},
\citealt{Dawson12}). The offsets of
different components of the subclusters reflect the differences in the
strengths of interactions between various components. Galaxies are
expected to lead the gas due to their negligible interaction cross
sections with other components. The intracluster medium (ICM) is expected to lose
momentum through electromagnetic interactions. On the other hand, offsets
between dark matter and galaxies may suggest dark matter self-interaction
(\citealt{Kahlhoefer14}, \citealt{Randall2008d}).  
\par
The galaxy cluster ACT-CL J0102-4915, (nicknamed ``El Gordo", at z=0.87),
was discovered via its Sunyaev-Zel'dovich (SZ) effect by the Atacama Cosmology Telescope (ACT;
\citealt{Menanteau2010}; \citealt{Marriage11}); it has the strongest SZ
effect of the full ACT survey \citep{Hasselfield2013}, and was discovered
to be undergoing a major merger approximately in the plane of the sky
(\citealt{M12}, hereafter M12). El Gordo possesses a range of noteworthy features that allow us to constrain the merger dynamics in multiple ways. 
\begin{figure*}
	\includegraphics[width=1\textwidth]{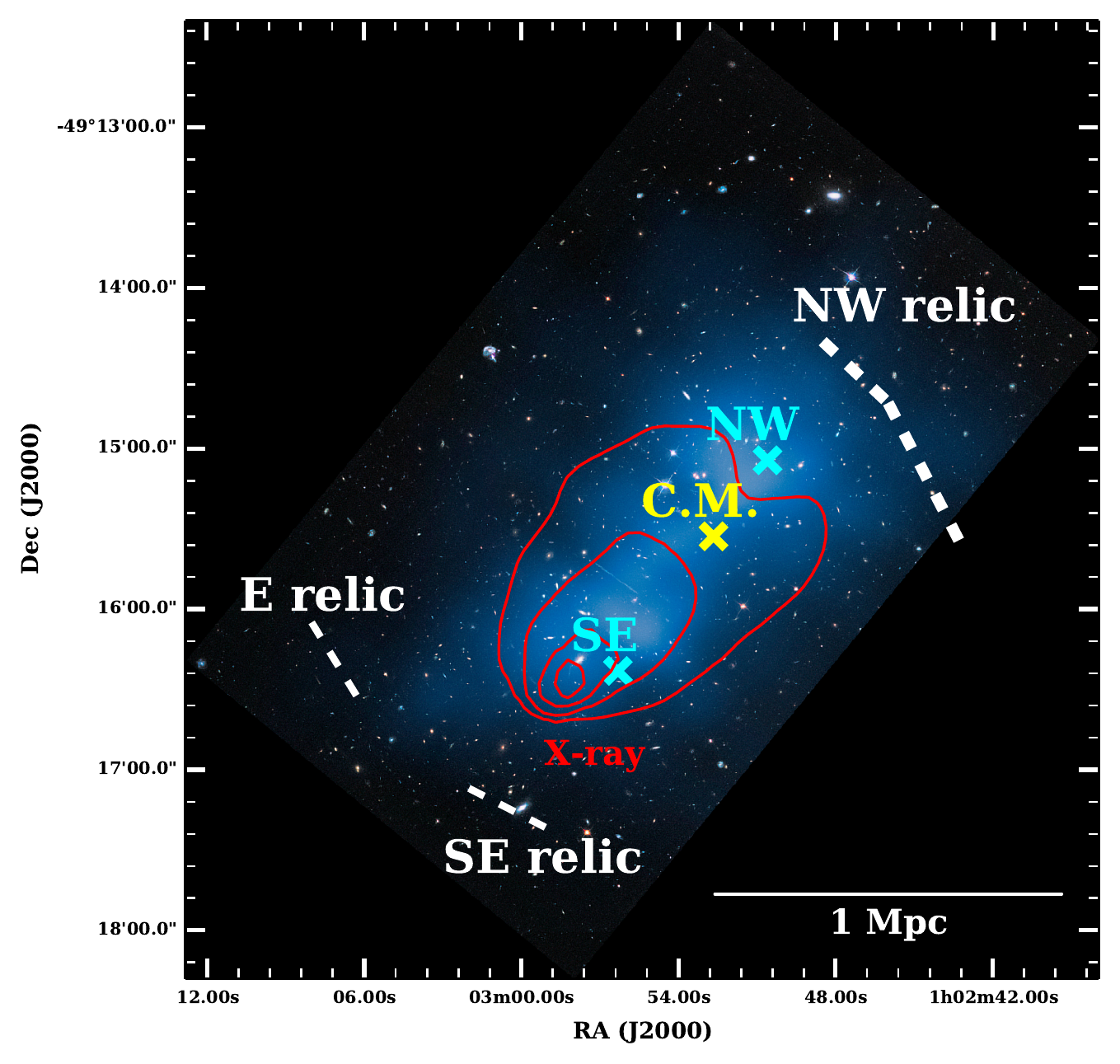}
	\caption{Configuration of El Gordo showing overlay of dark
		matter distribution in blue, and X-ray emission in red. 
		(Image credit: NASA, ESA and \citealt{Jee13}). 
		The cross markers show the positions of the northwest (NW) and
		southeast (SE) dark matter density peaks, and the center of mass (CM)
		locations respectively. Note that the mass ratio of the NW subcluster
		to the SE subcluster is $\sim 2:1$ \citep{Jee13}. 
		The dashed white lines indicate the approximate location and extent of the northwest radio relic (NW relic), the east radio relic (E relic) and the
		southeast radio relic (SE relic) \citep{L13}.
		\label{fig:config}
	}
\end{figure*}
From the spectroscopy and Dressler-Schectman test for the member galaxies
in \cite{Sifon13}, it is shown that El Gordo does not have complicated
substructures in its galaxy velocity distribution. 
El Gordo is further confirmed to be a binary merger 
from the weak lensing analysis by \cite{Jee13}. The weak lensing analysis shows
a mass ratio of $\sim$2:1  between the two main subclusters, named
according to their location as the northwest (NW) and southeast (SE) subclusters respectively 
(See Figure~\ref{fig:config}). A bimodal distribution of cluster member galaxies is
also observed \citep{M12}. In addition, El Gordo has an interesting X-ray morphology. In the northwest, it shows a wake feature, i.e.,
turbulent flow due to object of higher density moving through fluids, while
in the southeast, it shows the highest X-ray emissivity indicative of a
cool gas core at the head of the wake. The most straightforward explanation for this morphology is that the
cool gas core has passed from the northwest to the southeast
\citep{M12}. 
The high mass of El Gordo also makes it a good
gravitational lens. \cite{Zitrin13} have found multiple strong
gravitationally lensed images around the center region of El Gordo. 
On the outskirts, strong radio emission is detected in
the NW and the SE respectively. These radio emitting regions show steep spectral
index gradients and are identified as radio relics associated with shock waves
created from the merger \citep{L13}. El Gordo is one of $\sim 50$ galaxy clusters that have
been associated with a radio relic and show dissociation between the X-ray
gas and the DM subclusters. \par 

\begin{figure*}
	\includegraphics[width=\linewidth]{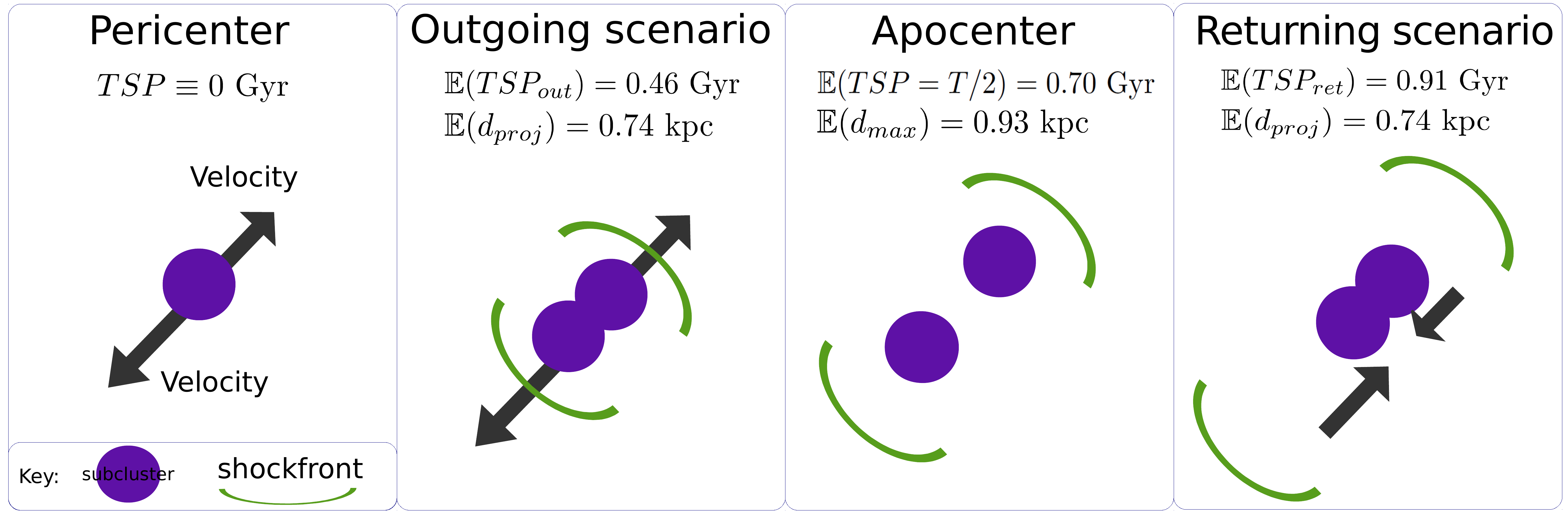}
	\caption{Illustration of the spatial location of different components of El Gordo at
		different stages of the merger. Earlier stages (with a smaller TSP) are on the left side of later stages. The rightmost returning scenario is preferred from our simulation.} 
	\label{fig:merger_cartoon}
\end{figure*}
In this paper, we combined most of the available information of El Gordo
with the main goal of giving estimates of
the dynamical parameters after taking into account all
constraints and uncertainties due to the missing variables.
Since mergers of clusters proceed on the time-scale of many millions of
years, one of the most important missing variables to infer is the
time-since-pericenter (TSP)$^\dagger$, which is defined to be the time when the mass
peaks of the DM subclusters are at minimum separation. \footnote{TSP in this
	paper is completely identical to the variable time-since-collision (TSC) in
	\cite{Dawson12}. We have renamed the variable to avoid confusion about how we
define collision as pericenter.}
Determining the TSP of similar clusters helps
us reconstruct different stages and recover the physics of a cluster merger.
In particular there is a degeneracy between the following two possible
scenarios:
We call the scenario for which the subclusters are
moving apart after pericenter to be ``outgoing", and the alternative scenario 
``returning" for which the subclusters are approaching each other after turning
around from the apocenter for the first time (See
Fig.~\ref{fig:merger_cartoon}).\par
Another crucial, missing piece of information is the 3D
configuration, i.e.\ the angle between the plane of the sky and the merger
axis called the projection angle $\alpha$. Since most of the dynamical
observables are projected quantities while the modelling of physics
requires 3D
variables, the deprojection contributes the
largest amount of uncertainty to the dynamical variables
(\citealt{D13}, hereafter D13). The morphology of the double relic of El Gordo suggests that
$\alpha$ should be small. 
For mergers with a
large projection angle, the radio emission would be projected towards the
center of the merger, which inhibits detection \citep{Vazza11}.
However, the only quantitative constraint on $\alpha$ for El Gordo is from
the analysis of the radio relic from \cite{L13} with a lower bound of $\alpha \geq 11.6 \degree$. A tighter
constraint on $\alpha$ is needed for us to reduce uncertainty of the
dynamical variables. 
\par 
We employed a data-driven approach that thoroughly probes parameter
space by directly drawing samples from the probability density functions
(PDFs) of
the observables. 
This work based on Monte Carlo simulation is particularly important since
the phase space of possible merger scenarios is large. Previous attempts at modeling El Gordo with hydrodynamical
simulations such as \cite{Donnert13} and \cite{Molnar14} provided only in
total a dozen possible configurations of El Gordo, which do not
reflect the range of input uncertainties. Another approach for
estimating dynamical parameters would be to look for multiple analogs of El Gordo in cosmological
simulations.  However, under the hierarchical picture
of structure formation in the $\Lambda$CDM model, there is a rare chance
for massive clusters like El Gordo to have formed at a redshift of $z = 0.87$.  
The number density of analogs with mass comparable to El Gordo in a
cosmological simulation is as low as $10^{-11} \mega\text{pc}^{-3}$
\citepalias{M12}.\par
In the following sections, we adopt the following conventions: (1) we
assume the standard $\Lambda$CDM cosmology with $\Omega_{m} = 0.3$,
$\Omega_{\Lambda} = 0.7$ and $H_0 = 70~\kilo\meter / \second / \mega
{\rm pc}$. (2) All confidence intervals are quoted at the 68\% level unless otherwise stated. 
(3) All quoted masses ($M_{200c}$) are based on mass contained
within $r_{200}$ where the mass density is 200 times the critical density
of the universe at the cluster redshift of $z = 0.87$. 
\begin{figure}
	\includegraphics[width = \linewidth]{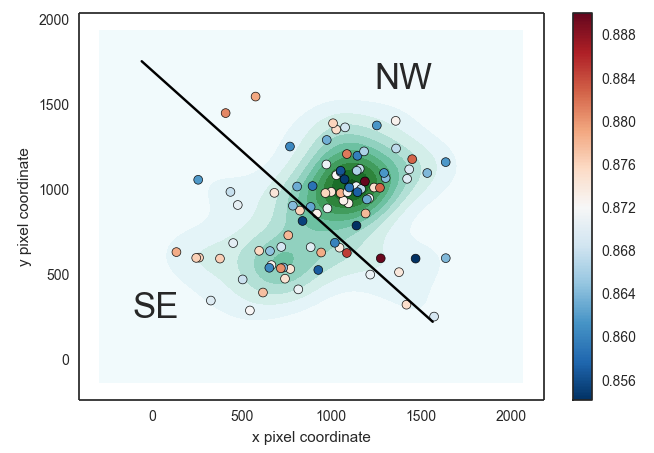}
	\caption{\label{fig:membership} Points showing the locations of the
	member galaxies and the division of the member galaxies among the two subclusters of El Gordo by a spatial cut
(black line). The color of the points shows the corresponding spectroscopic
redshift of the member galaxies (see color bar for matching of
spectroscopic values), with the redder end indicating higher
redshift. The galaxy number density contours in the background in green indicate a bimodal
distribution.} 
\end{figure}
\section{DATA} 
We gathered and analyzed data from multiple sources. 
See Table~\ref{tab:inputs} for descriptions of the PDFs of the input
variables. 
We examined the
spectroscopic data obtained from the Very Large Telescope (VLT) as described in \citetalias{M12} and \citet{Sifon13} for estimating the
relative velocity differences between the subclusters.
We adopted the identification of galaxy membership of El Gordo given by
\citet{Sifon13} with a total count of 89 galaxies.
To further distinguish member galaxies of each subcluster, we adopted the
spatial cut from \citetalias{M12}.
The adopted spatial cut is approximately perpendicular to the 2D merger
axis \citepalias{M12} and is consistent with
the bimodal number density contours (See Fig.~\ref{fig:membership}). 
There are 51 members identified in the NW subcluster and 35 members in the SE
subcluster. \par 
For the weak-lensing mass estimation, we used the
Monte Carlo Markov Chains (MCMC) mass estimates from \citetalias{Jee13}.
The Hubble Space Telescope data used in \citetalias{Jee13} were obtained from two
programs PROP 12755 and PROP 12477. PROP 12755 consisted of two pointings in the F625W, F775W, and F850LP.
F850LP for a 6' x 3' strip, while PROP 12477 provided a $2 \times 2$ mosaic
pattern with F606W and a single pointing in F814W. 
\par 
In order to further constrain our parameter space, we referred to the properties of
the radio relics from \citet{L13}. El Gordo shows radio emission on the
periphery of both subclusters \citepalias{M12}. The two radio relics, the
northwest (NW) relic and the southeast (SE) relic, of El Gordo were
tentatively identified
in the Sydney University Molonglo Sky Survey (SUMSS) data in low
resolution at 843 MHz \citep{Mauch03} as shown in M12. Higher
resolution radio observations conducted by \cite{L13} at 610 \mega Hz and
2.1 \giga Hz later confirmed the identities of the NW and the SE relic, and
found another extended source of radio relic in the east (E) (See Fig.~\ref{fig:config}). Among the radio relics, the NW relic possesses the most extended geometry
(0.56 Mpc in length), and its physics, including the
polarization and Mach number, were studied in the greatest detail. Such
information allows us to constrain $\alpha$ and the merger scenario. The E relic
was also reported to have a resolved length of 0.27 Mpc, while the SE relic
was found to overlap with a point source \citep{L13}. Both the E and SE
relic are closer to the SE DM subcluster, so we considered them to
originate from the same merger shock in the following work.
\section{METHOD -- Monte Carlo simulation} 
\label{sec:method}
We used the collisionless 
dark-matter-only Monte Carlo modeling code written by \citetalias{D13}, to
model the dynamics of the DM subclusters of their first core-passage.
In the D13 code, the time evolution of the
head-on merger was computed based on an analytical, dissipationless model
assuming that the only dominant force is the gravitational attraction from
the masses of two Navarro-Frenk-White (hereafter NFW) DM halos
\citep{Navarro96}. 
The gravitational attraction was evaluated and summed at 10 000 fixed grid
points of each of the analytic NFW halo profiles out to the respective
$r_{200c}$.\par
In the simulation, many realizations of the collision are
computed by drawing random realizations of the PDFs of the inputs. Most
input variables are obtained from previous observations ($\vec{D}$).  One
unknown model variable, which is the projection angle between the plane of the sky
and the merger axis, $\alpha$, is drawn from the PDF of $\alpha$ being
observed: 
\begin{equation}
	\alpha^{(j)} \sim f(\alpha) = \cos \alpha.
\end{equation}
and the calculation of the output variables of the $j$-th realization can be denoted as: 
\begin{equation}
	(\vec{\theta}^\prime)^{(j)} = g(\alpha^{(j)}, \vec{D}^{(j)}), 
\end{equation}    
for a suitable function $g$ that describes conservation of energy during
the collision of the two NFW halos due to the mutual gravitational
attraction.  In particular, the required $\vec{D}$ includes the masses ($M_{200_{NW}},M_{200_{SE}}$) the redshifts ($z_{NW}, z_{SE}$) and the
projected separation of the two subclusters ($d_{\rm proj}$).  See
Table~\ref{tab:inputs} for quantitative descriptions of the sample PDFs, and
the outputs with physical importance are described in detail in Section~\ref{sec:outputs}. \par
Finally, we excluded realizations that produce any unphysical output
values, such as realizations with time-since-pericenter larger than the age of universe at the
cluster redshift.  We refer to this process of excluding unphysical
realizations as applying weights. 
To ensure convergence of the output PDFs, in total, 2 million realizations
were computed. However, the estimates would agree up to 1\% just 
from 20 000 runs \citepalias{D13}. 
Even though we describe the weights for one variable at a time 
(See Appendix~\ref{app:priors}), 
the correlations between different variables are properly taken into account
since we discarded all the variables of the problematic
realizations.\par 
\begin{table}
	\caption{Properties of the input sampling PDFs ($\vec{D}$) of the Monte Carlo
simulation. We obtained estimates of the inputs via different methods. $^a$We
made used of the MCMC chains from \citetalias{Jee13} for mass-related
estimates (See Section \ref{subsubsec:WL_mass_estimate}
). $^b$The redshift distributions were obtained from bootstrapping (See Section 
\ref{subsubsec:membership_and_redshift}).
$^c$We approximated the positions of the centroids with 2D Gaussians before we
calculated the projected separations of the subclusters (See Section 
\ref{subsubsec:proj_sep}). Even though the distributions of the mass estimates 
and the redshift were not estimated via parametric methods (e.g. fitting mean
and variances of Gaussians), they
resemble Gaussian distributions due to the Central Limit Theorem.
} 
\begin{center} 
\begin{tabular}{@{}lccccc}
\hline \hline Data & Units & Location & Scale & Ref \\ \hline
$M_{200c_{\mathrm{NW}}}$ & $10^{14} h_{70}^{-1}$ M$_{\odot}$ &13.0&1.6&
\citetalias{Jee13}$^a$\\ 
c$_{\mathrm{NW}}$ &  & 2.50& 0.02& \citetalias{{Jee13}}$^a$ \\ 
$M_{200c_{\mathrm{SE}}}$ & $10^{14} h_{70}^{-1}$ M$_{\odot}$ &7.6&1.2 &
\citetalias{Jee13}$^a$\\ 
$c_{\mathrm{SE}}$ &  & 2.70 & 0.04& \citetalias{Jee13}$^a$\\ 
$z_{\mathrm{NW}}$ &  & 0.86842 & 0.00109& \citetalias{M12}$^b$\\ 
$z_{\mathrm{SE}}$ &  & 0.87110 & 0.00117& \citetalias{M12}$^b$\\ 
d$_{\mathrm{proj}}$ & Mpc & 0.74 &0.007 & \citetalias{Jee13}$^c$\\ 
\hline 
\end{tabular} 
\end{center} 
\label{tab:inputs} 
\end{table} 
The system of El Gordo satisfies several major assumptions in the Monte Carlo
simulation. One of the strongest assumptions is that all realizations correspond to
gravitationally bound systems. The simulation excludes all realizations
that would result in relative pericenter velocities of the subclusters
higher than the free-fall velocity. We justify our assumption of
modeling only gravitationally bound systems by noting that the relative escape
velocity of the subclusters for El Gordo is
$4500~\kilo\meter~\second^{-1}$ (based on the mass estimates of
\cite{Jee13}). Studies from cosmological simulations giving the PDFs of the pairwise
velocities of massive merging clusters ($>10^{15} M_{\sun}$) indicate that it
is highly unlikely that the pairwise velocities would be $> 3000~\kilo
\meter~\second^{-1}$ under $\Lambda$CDM (\citealt{Thompson12},
\citealt{Lee2010}).\par
Other major assumptions for modeling systems with this code include negligible impact parameter.
Several papers have noted that the X-ray morphology of a bimodal merger is
sensitive to the impact parameter (\citealt{Springel2007},
\citealt{Ricker98}, \citealt{Mastropietro2008a}); an impact
parameter as small as 0.1 Mpc can result in substantial asymmetry. 
The X-ray morphology of El Gordo is approximately symmetric about the merger axis. 
On the other hand, the dynamics of the merger is not as sensitive to the
impact parameter as the X-ray morphology. The simulations of
\cite{Ricker98} of bimodal mergers of $10^{15} M_{\sun}$ halos, showed that the
resulting relative velocity would be approximately $2000 \text{ km s}^{-1}$, relatively
insensitive to impact parameters between 0 to 5 times the scale radius ($5~r_s = 3.5$ Mpc). \cite{Mastropietro2008a} also reported that
an impact parameter of $0.1~r_{200} = 0.14$ Mpc affected merger dynamics only
at the $\sim$10\% level. \citealt{Molnar14} indicated that the impact parameter of El Gordo
may be as large as $40\%~r_s \approx 0.3$ Mpc, where $r_s$ is
the corresponding characteristic core
radius of the NFW halo with the mass of the SE subcluster. We attribute
the result from \citealt{Molnar14} to incomplete exploration of
the parameter space, and note that other impact parameter values may also match the
X-ray observables of El Gordo. \par
Other assumptions in this simulation include negligible dynamical friction
during the merger, negligible mass accretion and negligible self-interaction
of dark matter. Discussion of the effects of each of these assumptions is
included in \citetalias{D13}.  
\subsection{Inputs of the Monte Carlo simulation} \label{sec:inputs}
\subsubsection{Membership selection and redshift estimation of subclusters}
\label{subsubsec:membership_and_redshift}
After identifying members of each subcluster, 
we performed 10, 000 bootstrap realizations to estimate the biweight
locations of the spectroscopic redshifts of the respective members in order
to obtain the samples of the PDFs of the redshifts of each subcluster. 
The spectroscopic redshift of the subclusters were
determined to be 
$z_{\mathrm{NW}} = 0.86842 \pm 0.0011$ and 
$z_{\mathrm{SE}} = 0.87131 \pm 0.0012$, where the quoted numbers represent the
biweight location and 1$\sigma$ bias-corrected confidence level
respectively \citep{Beers90}.  
Both the estimated redshifts of the subclusters and the uncertainties are
consistent with estimates of $z=0.8701 \pm 0.0009$ for El Gordo given by 
\citealt{Sifon13}, and the fact that the
member galaxies of El
Gordo shows large velocity dispersion, i.e.\ the largest velocity
dispersion among all the ACT galaxy clusters, as reported by
\citetalias{M12}.
\begin{figure}
	\includegraphics[width = \linewidth]{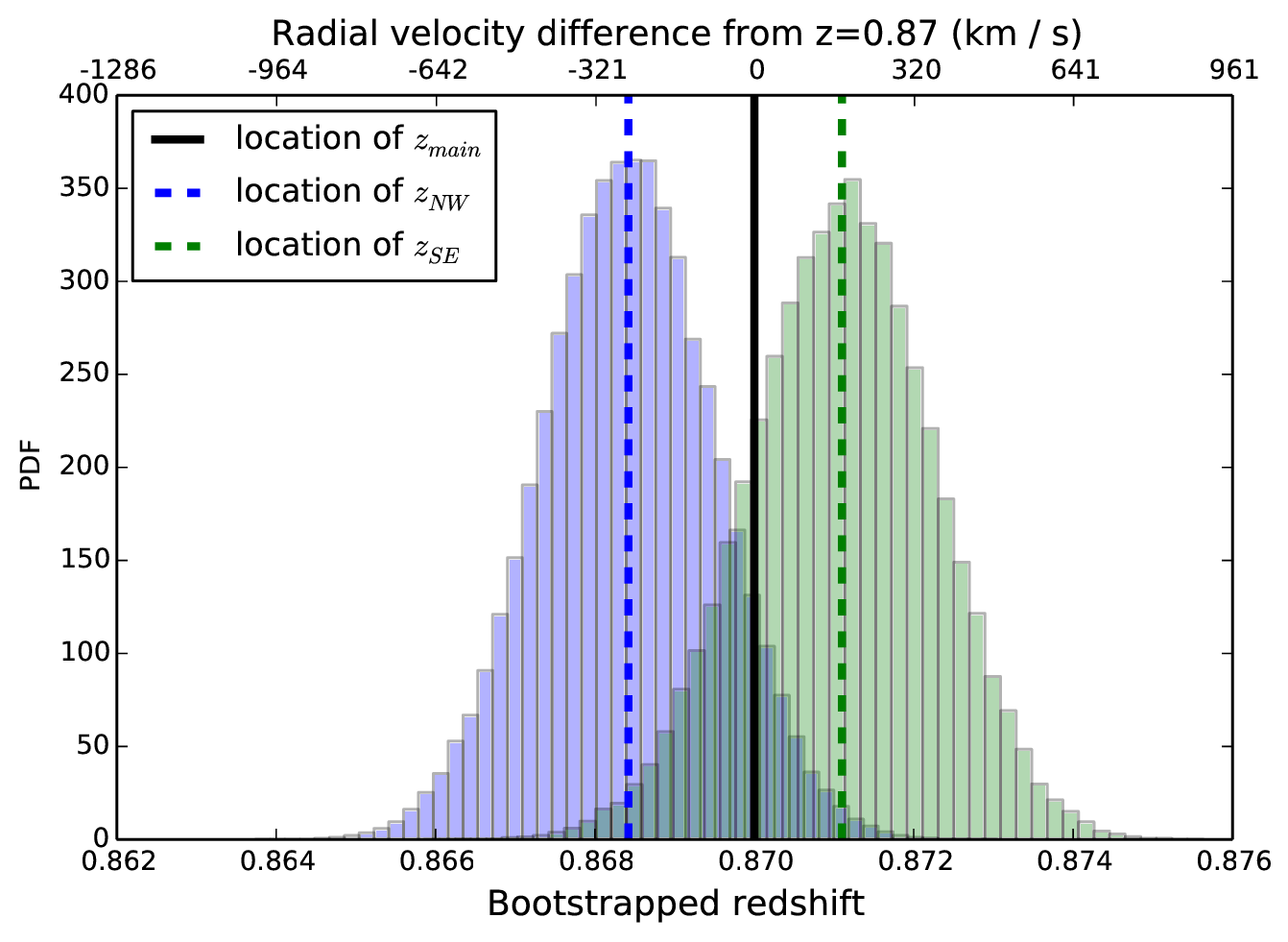}
	\caption{Bootstrapped location of the
	redshift estimates and $v_{rad}$ estimates for each subcluster using the
	selected spectroscopic members. The shaded histograms represent the
	bootstrapped samples.
} \label{fig:bootstrap_redshift}
\end{figure}
We estimated the radial velocity differences of the
subclusters by first calculating the velocity of each subcluster with
respect to us, using  
\begin{equation}
	v_i = \left[ \frac{(1+z_i)^2 - 1 }{(1+z_i)^2 + 1 }\right]c,
\end{equation}
where $i=1, 2$ represents the two subclusters, and $c$ is the speed of
light. The relative radial velocity was calculated by: 
\begin{equation}
	\Delta v_{rad}(t_{obs}) = \frac{|v_2 - v_1|}{1-\frac{v_1 v_2}{c^2}}.
\end{equation}
We obtained a low radial velocity difference of the two subclusters to be
$476~\pm~242~\kilo\meter~\second^{-1}$ (See Fig.~\ref{fig:bootstrap_redshift}). 
The radial velocity difference of $586~\kilo \meter~\second^{-1}$ reported by \citetalias{M12} 
is higher than our estimates but within the 68\% bias-corrected
confidence interval.\footnote{bias from the location estimator of
	the bootstrapped distribution not giving the maximum likelihood value was
corrected for.} 
% Limitations and possible improvements of this analysis
% of $v_{rad}$ are provided in the discussion. 
\subsubsection{Weak lensing mass estimation} 
\label{subsubsec:WL_mass_estimate}
We obtained 40, 000 samples of the joint PDFs of the masses of the two dark
matter halos as the outputs of the Monte Carlo Markov Chain (MCMC)
procedure from \citealt{Jee13}. Discussion of the handling of the weak
lensing source galaxies and the details of the MCMC procedure for mass
estimation can be found in \citealt{Jee13}. 
\subsubsection{Estimation of projected separation ($d_{\rm proj}$)} 
\label{subsubsec:proj_sep}
To be consistent with our MCMC mass inference, our Monte Carlo simulation takes 
the projected separation of the NFW halos to be those of the inferred
DM centroid locations in \citealt{Jee13}. We drew random samples
 of the location of centroids from two 2D Gaussians centered at
 R.A.$=$01:02:50.601, Decl.=$-$49:15:04.48 for the NW subcluster and R.A. =
 01:02:56.312, Decl.=$-$49:16:23.15 for the SE
subcluster, with a 1'' standard deviation each as estimated from the
convergence map of \citet{Jee13}. The
inferred centroid locations correspond to a mean projected separation
($d_{\rm proj}$) of $0.74\pm {0.007}$ Mpc.  
\subsection{Outputs of the Monte Carlo simulation}\label{sec:outputs}
We outline the outputs of the simulation here to facilitate the discussion
of the design of the weights used in the simulation. The simulation
provides PDF estimates for 8 output variables. Variables
of the highest interest include the time-since-pericenter and the angle $\alpha$, which is
defined to be the projection angle between the plane of the sky and the
merger axis. Other output variables are dependent on $\alpha$ and time. Specifically, the simulation denotes the time dependence by
providing several characteristic time-scales, including the time
elapsed between consecutive collisions
($T$) and the time-since-pericenter of the observed state ($TSP$), with the
time of pericenter defined to be when the centers of the two NFW halos coincide. 
\par
We provide two versions of the time-since-pericenter variables $TSP_{\rm out}$ and
$TSP_{\rm ret}$ to denote different possible merger scenarios. 
1) The TSP for the``outgoing" scenario corresponds to the
smaller $TSP_{\rm out}$ value, and 2) the ``returning'' scenario 
corresponds to the larger $TSP_{\rm ret}$.
We describe how we make use of properties of the radio relic to evaluate
which scenario is more likely in
section~\ref{sec:positionprior}. Evolution of the merger after the second
passage is not considered. Outputs from our dissipationless simulation for
a ``second'' passage will not differ from the first passage, and the
predicted relic position would be so far for us to rule this out.
 
The simulation also outputs estimates of variables that describe
the dynamics and the characteristic distances of the merger. The relative
3D velocities of the subclusters, both at the time of the
pericenter ($v_{3D}(t_{\rm per})$) and at the time of observation
($v_{3D}(t_{obs})$) are provided. The characteristic
distances included in the outputs are the maximum 3D separation ($d_{max}$),
which is the distance between the subclusters at
the apocenter and the 3D separation of the subclusters at observation
($d_{3D}$). 
\subsection{Design and application of weights} 
\label{sec:priors}
One of the strengths of the Monte Carlo simulation by
\citetalias{D13} is its ability to detect and rule out extreme input values that would result in
unphysical realizations via the application of weights. 
Our default weights are described in D13 and we include them in
Appendix~\ref{app:results} for the convenience of the readers. 
In addition, we have devised a new type of weights of the projection angle $\alpha$
based on the polarization fraction of the radio relic.

%---------------------------------------------------------------------------

\subsubsection{Monte Carlo weights based on the polarization fraction of the radio relic}
\label{subsubsec:polar_frac}

We can relate the polarization fraction of the radio relic to the
projection angle by examining the
generating mechanism of the radio relic.
The observed radio relic was generated by synchrotron emission of free electrons in a
magnetic field. If the magnetic field was uniform, the observed
polarization fraction of the synchrotron emission of the electrons depends on the
viewing angle (or equivalently the projection angle) with respect to the
alignment of the magnetic field. Synchrotron emission from electrons inside
unorganized magnetic field is
randomly polarized. The high reported integrated polarization fraction from
\citet{L13} can be explained by a highly aligned magnetic field,
compressed along with the ICM during a merger
(\citealt{E98}, \citealt{vanWeeren10}, \citealt{Feretti12}).
%This picture is consistent with a high polarization fraction.
\par

We designed the weights to reflect how $\alpha$ decreases monotonically as the
maximum observable integrated polarization fraction ($\langle P \rangle$). 
\begin{figure}
	\includegraphics[width=\linewidth]{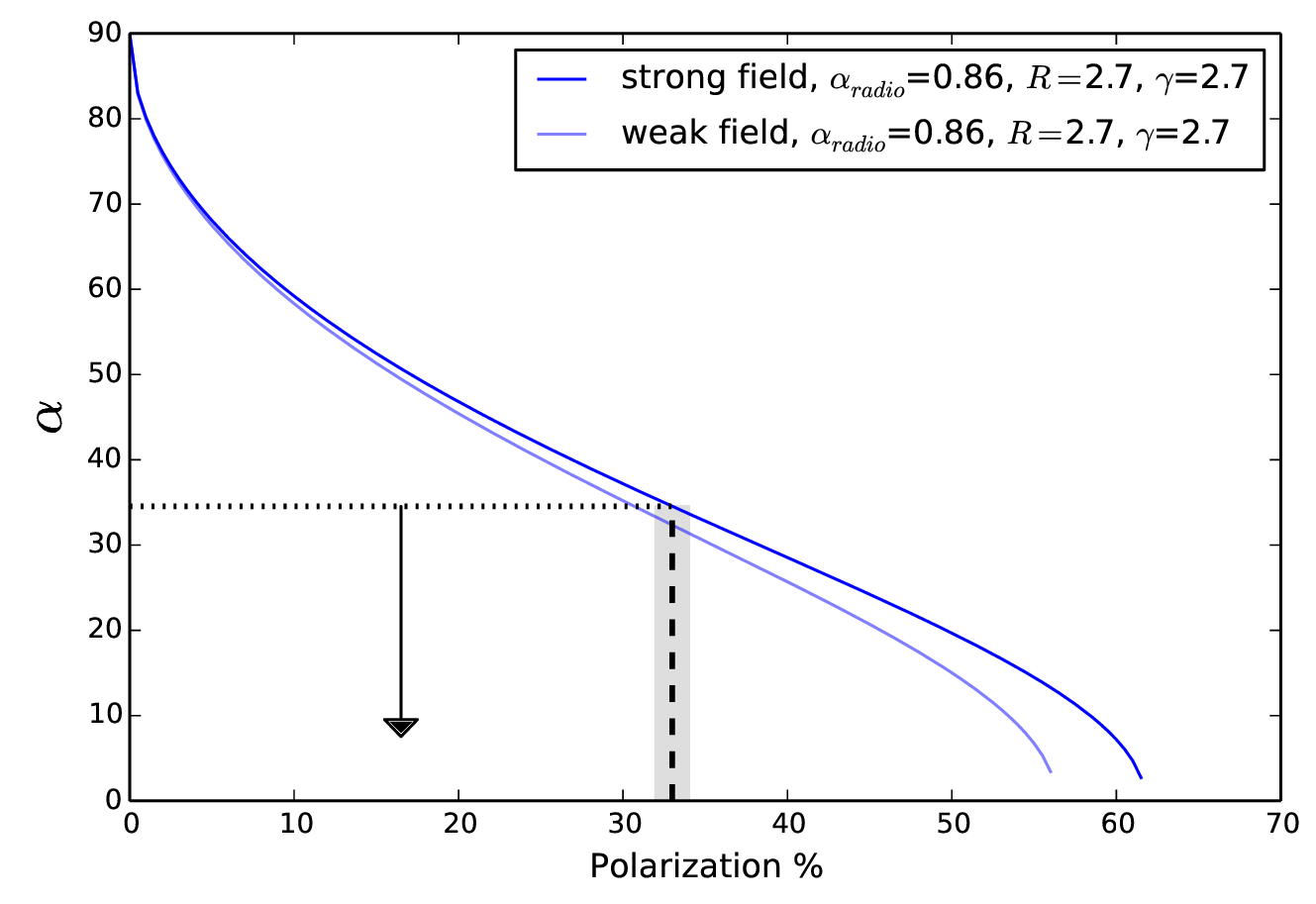}
	\caption{Predictions of polarization percentage of the radio relic at a
		given projection angle from different models, reproduced from
		\citep{E98} or equation~\ref{eqn:Ensslin_model}. Each model assumes electrons producing the radio emission
		to be accelerated inside uniform magnetic field of various strengths ({\it strong} or 
		{\it weak}). The curves are plotted with spectral index of the radio emission
		($\alpha_{radio}$), spectral index of the electrons ($\gamma$) and
		compression ratio of the magnetic field ($R$) corresponding to the
		estimated values from \citet{L13}.
		We highlight the observed polarization percentage of the main NW radio relic
		of El Gordo by the dotted vertical line with the greyed out region
		indicating the uncertainty \citep{L13}.\label{fig:Ensslin_fig}}
\end{figure}
This assumption is based on the class of models given by \cite{E98}(See
Figure~\ref{fig:Ensslin_fig}). In particular, we refer to a model from \cite{E98} 
that would give the most
conservative estimate on the upper bound of $\alpha$:
\begin{equation}
	\alpha = 90 \degree - \arcsin \left( \sqrt{\frac{\frac{2}{15} \frac{13R -
	7}{R - 1} \frac{\gamma + 7/3}{\gamma + 1} \langle P_{strong} \rangle}{1 +
		\frac{\gamma + 7/3}{\gamma +1} \langle P_{strong} \rangle
	}}\right),\label{eqn:Ensslin_model}
\end{equation}
This model corresponds to the case of a strong field with the relic being supported by
magnetic pressure only, with the spectral index of the radio
emission being $\alpha_{radio} = 0.86$, the compression ratio of the
magnetic field being
$R=2.7$ and the spectral index of the electrons being $\gamma = 2.7$
\citep{L13}. 
This model predicts a maximum integrated polarization fraction 
 of
$\sim60\%$ when $\alpha \rightarrow 0$. 
%We consider 39\degree as an upper bound on the projection angle since this 
% idealized model assume isotropic distribution of magnetic field and
%electrons. 
This  polarization fraction of $\sim60\%$ predicted by \citep{E98} is
consistent with the upper bound of relic polarization fraction in cosmological
simulations \citep{S13}. From this model, the
observed integrated polarization fraction of $33\%\pm1\%$ corresponds to an estimated value
of $\alpha  = 35\degree$. 
No other model of the magnetic field should predict 
a higher polarization fraction, thus it is highly unlikely that we see 33\%
integrated polarization at $\alpha > 35\degree$.  
\par

We cannot rule out $\alpha \leq 35\degree$ because magnetic field
nonuniformities can lower the polarization below the Ensslin model value.
\cite{E98} assumes an isotropic distribution of electrons in an isotropic magnetic field. Cosmological
simulations of radio relics from \cite{S13} show varying polarization
fraction across and along the relic assuming $\alpha = 0$, resulting in a
lower integrated polarization fraction. For example, it is possible to see 
an edge-on radio relic ($\alpha = 0$) with integrated polarization fraction of 33\%. 
Furthermore, \cite{S13} shows that after convolving the
simulated polarization signal with a Gaussian kernel of 4\arcmin~to
illustrate effects of non-zero beam size, the polarization fraction drops
to between 30\% to 65\% even when $\alpha = 0$. We examined the effects of perturbing
the cutoff value of this weight to ensure the uncertainties do not
introduce significant bias in the estimated output variables in
section~\ref{sec:sensitivityTests}.
To summarize, we used a conservative uniform weight to encapsulate the
information from the polarization fraction of the radio relic as:
\begin{equation}
w(\alpha) = 
	\begin{cases}
	& \text{const. for  }\alpha < 35 \degree \\ 
	& 0 \text{ otherwise,}\label{eqn:polarprior}.
	\end{cases}
\end{equation}
We refer to equation~\ref{eqn:polarprior} as the polarization weights. Unless
otherwise stated, the main results of the paper are obtained after applying
this polarization weight in addition to the default weights.

%%---------------------------------------------------------------------------

\subsection{Extension to the Monte Carlo simulation - Determining merger
scenario with radio relic position by model comparison}
\begin{table*}
	\begin{minipage}{170mm} 
	\caption{List of variables that provide quantitative constraints for the merger
		scenario. For details of the distribution of each variable see the corresponding
	$^\dagger$Section. Calculations were done with all available realizations
instead of the best estimate value listed here.}
\begin{center} 
\begin{tabular}{@{}lccccc}
	\hline \hline Variable & Best estimate value & Unit & Section$^{\dagger}$ \\ 
	\hline
Time averaged speed of SE relic in the CM frame ($\langle v_{relic} \rangle$) 
& 530 &  km~s$^{-1}$  & \ref{sec:merger_scenario}\\
Time averaged speed of NW relic in the CM frame ($\langle v_{relic} \rangle$) &
  310 &  km~s$^{-1} $ & \ref{sec:merger_scenario}\\
Projected separation of SE relic from the CM ($s_{proj}$)  & 1.1  & Mpc & \ref{sec:positionprior} \\ 
Projected separation of NW relic from the CM ($s_{proj}$)  & 0.63 & Mpc & \ref{sec:positionprior} \\ 
Outgoing time-since-pericenter ($TSP_{out}$)  &  0.61 & Gyr &
\ref{sec:outputs}, 
\ref{sec:merger_scenario}
\\ 
Returning time-since-pericenter ($TSP_{ret}$)  & 1.0 & Gyr & \ref{sec:outputs}, 
\ref{sec:merger_scenario}
\\ 
Age of the universe at z = 0.87 & 6.30 & Gyr&  \\ 
Free fall velocity of subclusters & 4500 & km~s$^{-1}$ & \ref{sec:method}\\  
Polarization fraction of NW relic ($\langle P_{strong} \rangle$) & 33\% &  &
\ref{subsubsec:polar_frac}\\ 
\hline 
\end{tabular} 
\end{center} 
\label{tab:input_contraints} 
\end{minipage}
\end{table*} 

One of the biggest questions involving the merger is whether El Gordo is
observed during a returning or outgoing phase. We compared the two merger
scenarios by making use of the observed projected separation of the relic from the
center of mass.
Simulations of cluster mergers such as the work of \citet{Paul2011b},
\citet{VanWeerenRJ2011}, and \citet{Springel2007} showed that merger shock
fronts that may correspond to the radio relics 1) are generated near the
center of mass of the subclusters close to the time of the first
core-passage, 2) propagate outward with the shock speed decreasing only slightly.
The propagation speed of the shock wave {\it with respect to the
center-of-mass} is reported 
to drop between $\sim 10\%$ from \citet{Springel2007} and  
10\% to 30\% from \citet{Paul2011b} (note that the   
 time resolution is 0.6 Gyr for the simulation in \citealt{Paul2011b}).
\par 
To capture the monotonically decreasing trend of the
propagation speed of the shock fronts with respect to the center of
mass, we expressed the possible time-averaged shock speeds as a factor of the inferred
pericenter speed of the corresponding subcluster in the center of mass
 frame. 
Then we calculated how far the shock would have propagated for our inferred
$TSP_{\rm out}$ and $TSP_{\rm ret}$ values. We worked in the center of mass frame where the
shock speed is expected to drop slightly with TSP. 
The projected separation of the shock is approximated as:
\begin{equation}
	%s_{proj} = \langle v_{relic} \rangle (\hat{t}_{obs} - \hat{t}_{col})
	s^j_{proj} \approx \langle v_{relic} \rangle^j (t^j_{obs} - t^j_{per})
	\cos(\alpha^j),
	\label{eq:proj_s_model}
\end{equation}
where the superscript $j$ of any variable denotes the value of the
variable from the j-th realization of the simulation, and $s_{proj}$ is the estimated projected
separation. We estimated the upper and lower bounds of the time-averaged velocity
$\langle v_{relic} \rangle$ of the shock between
the pericenter of the subclusters and the observed time as:  
\begin{align}
	\label{eqn:NW_speed}
	\langle v_{NW relic} \rangle^j &= \beta~v^j_{3D, NW}(t_{\rm per}) \\
	&= \beta~v^j_{3D}(t_{\rm per}) \frac{m^j_{SE}}{m^j_{SE} + m^j_{NW}}, 
\end{align}
where  $\beta$ is a factor that we introduce to represent the
uncertainty of the velocity of the relic shock wave, $v_{3D, NW}(t_{\rm per})$ 
refers to the pericenter velocity of
the NW subcluster in the center-of-mass frame as a comparison, and $m$
represents the mass within $r_{200c}$ of each subcluster as denoted by the
labels in the subscripts. 
Likewise, we have also computed the expected projected separation of the SE
relic using:  
\begin{equation}
	\label{eqn:SE_speed}
	\langle v_{SE relic} \rangle^j = \beta~v^j_{3D}(t_{\rm per}) \frac{m^j_{NW}}{m^j_{SE} + m^j_{NW}}. 
\end{equation}
\par 
\begin{figure}
	\includegraphics[width=\linewidth]{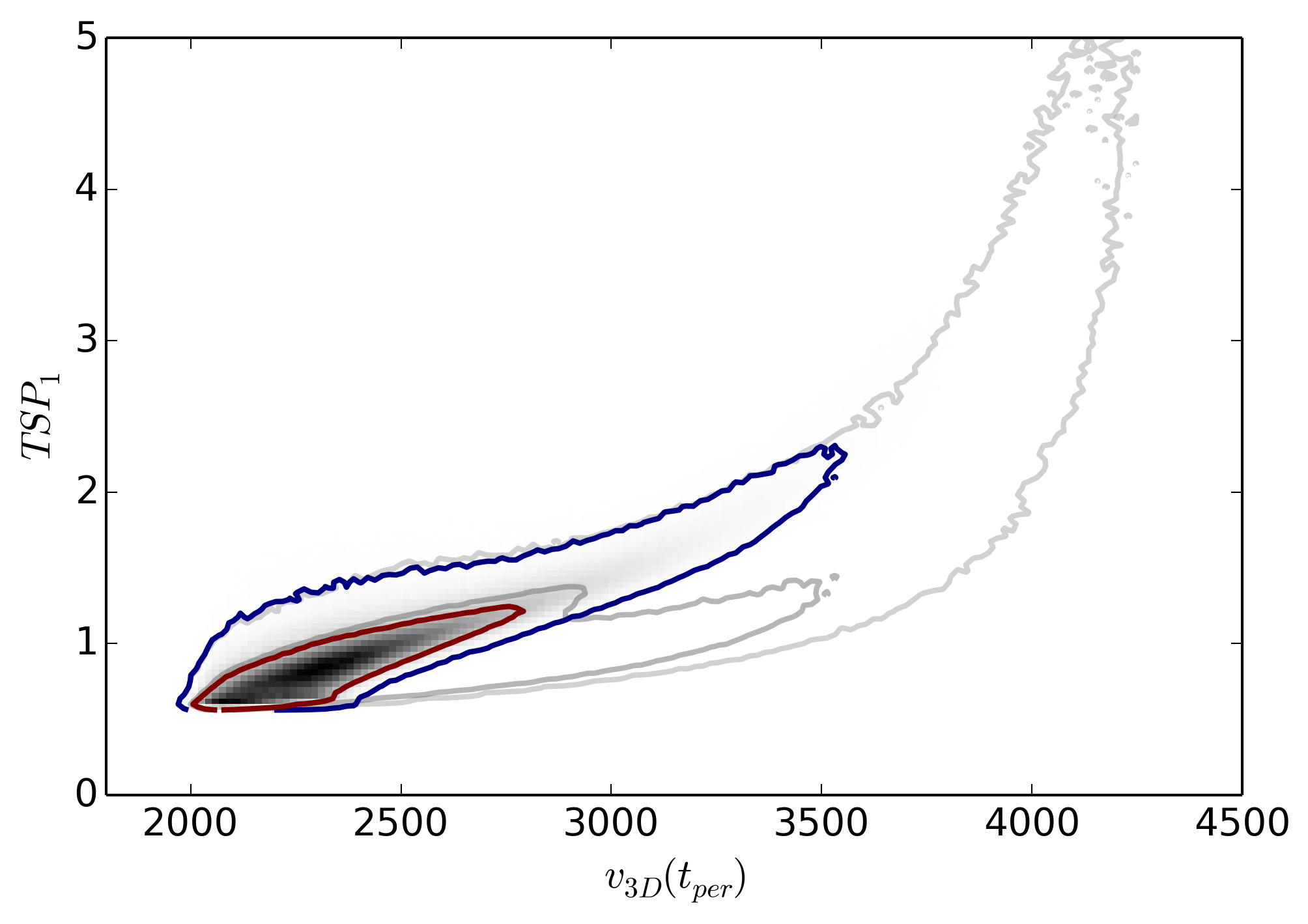}
	\caption{The marginalized output PDF of the returning time-since-pericenter
($TSP_{\rm ret}$) vs. the 3D velocity at the time of pericenter for El Gordo. The
grey set of contours show the confidence regions before applying the
polarization weight and the colored contours correspond to the confidence
regions after applying the weights. The contours represent the 95\% and
68\% confidence regions respectively. }
	\label{fig:TSP_v3D}
\end{figure}

\begin{table*}
\begin{minipage}{170mm} 
\caption{Table of the output PDF properties of the model variables and output variables from Monte Carlo simulation
\label{tab:outputs}}
\begin{tabularx}{\textwidth}{@{\extracolsep{\fill}}lccccccccc@{}}
\hline
\hline
&&&&Default weights & & & & Default + polarization weights  \\ 
\cmidrule{4-6} \cmidrule{8-10} 
Variables & Units && Location & 68$\%$ CI $^{\dagger}$ &95$\%$ CI && Location & 68$\%$ CI  & 95$\%$ CI \\ 
\hline 
$\alpha$ &(degree)&&43&19-69&6-80&&21&10-30&3-34\\
$d_{\rm proj}$ &Mpc&&0.74&0.74-0.75&0.73-0.76&&0.74&0.74-0.75&0.73-0.76\\
$d_{\rm max}$ &Mpc&&1.2&0.90-2.2&0.77-4.6&&0.93&0.81-1.2&0.75-1.9\\
$d_{\rm 3D}$ &Mpc&&1.0&0.79-2.1&0.75-4.3&&0.80&0.76-0.88&0.74-0.91\\
$TSP_{\rm out}$&Gyr&&0.61&0.4-0.95&0.26-2.4&&0.46&0.30-0.55&0.21-0.64\\
$TSP_{\rm ret}$&Gyr&&1.0&0.77-1.7&0.63-4.4&&0.91&0.69-1.3&0.59-2.3\\
$T$&Gyr&&1.6&1.3-2.4&1.2-6.4&&1.4&1.2-1.6&1.2-2.2\\
$v_{\rm 3D}(t_{\rm obs})$ & \kilo \meter~\second$^{-1}$ &&580&260-1200&59-2400&&940&360-1800&62-2900\\
$v_{\rm rad}(t_{\rm obs})$ & \kilo \meter~\second$^{-1}$ &&360&140-630&27-880&&310&110-590&8-840\\
$v_{\rm 3D}(t_{\rm per})$ & \kilo \meter~\second$^{-1}$ &&2800&2400-3700&2100-4200&&2400&2200-2800&2100-3500\\
\bottomrule
\end{tabularx}\\
\footnotesize{$\dagger$ CI stands for confidence interval}\\
\end{minipage}
\end{table*}

For the most likely range of $\beta$, we refer to the
simulations of cluster mergers by both \cite{Springel2007} and
\cite{Paul2011b} because those are some of the few simulations available that
quote shock propagation speeds in the center-of-mass frame of the cluster,
rather than the ICM frame. 
The simulation of the Bullet Cluster by \cite{Springel2007}, 
indicates that the propagation velocity of the shock evolves such that $\beta \approx 0.95$ within
$\sim 0.4~\giga$yr after the pericenter. For the analysis of El Gordo,
we suggest $\beta \approx 0.9$ to be the most likely value given that the $TSP$ of
El Gordo is longer. 
To include the possible range of valid $\beta$ values,
we examined $0.7 \leq \beta \leq 1.5$. This range of $\beta \approx 1$ allows us to use  
equations~\ref{eqn:NW_speed} and~\ref{eqn:SE_speed} to reflect that the
shock is driven by the merger. We note that the propagation speed of the
shock is also determined by the temperature, density and other details of
the gas medium (\citealt{Prokhorov2007}, \citealt{Springel2007},
\citealt{Milosavljevic07}), so it is physically possible for the shock to
propagate with $\beta > 1$. An example of
merging clusters (Merger F) with $\beta \approx 2.0$ has been reported by the cosmological
simulations in \citealt{Paul2011b}. 
However, we note that Merger F from \citealt{Paul2011b} has a lower mass of
$\sim 1.1 \times 10^{14} M_{\sun}$ in total, and that the coarse time resolution
in \citealt{Paul2011b} likely underestimates the
pericenter velocity and overestimates $\beta$.
We therefore
suggest a most likely range, closer to the value of $\beta$ inferred from
\cite{Springel2007}, as $0.7 \leq \beta \leq 1.5$.
In section~\ref{sec:merger_scenario}, we demonstrate that $\beta$ has to be
larger than $1.5$ to 
avoid the returning model, and in Appendix
~\ref{app:Bayes_factor}, we show the full range of possibilities up to
$\beta = 2.0$. 
In fig.~\ref{fig:our_guessed_scenario} and~\ref{fig:our_guessed_scenario1}, we compared our estimates of the projected locations of the relics to the
observed location given by Table 3 of L13. The given NW relic and E relic
locations are R.A.=01:02:46, Decl.=$-$49:14:43 and R.A.=01:03:07,
Decl.=$-$49:16:16
respectively. These locations correspond to a projected separation of 0.63 Mpc
(NW relic) and 1.1 Mpc (E relic) from the center of mass.

\label{sec:positionprior}

\section{RESULTS} 
We present an
overview of all the estimated variables in Table~\ref{tab:outputs}, with
results only applying the default weights on the left hand side of the table
and those also applied with the polarization weight on the right hand side.
Furthermore, we include the plots of all the marginalized PDFs with the
polarization weight in Appendix~\ref{app:results}. \par
We found that the two subclusters collided with a relative velocity of $2400\pm^{900}_{400}~\kilo\meter~\second^{-1}$, at an estimated projection
angle of $\alpha = 21\degree\pm^{9}_{11}$. From our analysis of the two
scenarios, we found that El Gordo is more likely to be observed at a returning
phase with an estimate of $TSP_{\rm ret} = 0.91\pm^{0.22}_{0.39}$ Gyr
(See section~\ref{sec:merger_scenario} and Appendix~\ref{app:Bayes_factor}
for a full discussion of the assumed relic propagation speed). This
returning scenario puts the
estimate of the time of pericenter to be when the age of the universe was
$\sim5.4$ Gyr. 
Our estimate of $v_{3D}(t_{obs})$ is
$940\pm^{860}_{580}~\kilo\meter~\second^{-1}$. (See Fig.~\ref{fig:TSP_v3D})
%is compatible with the independent estimate from \citealt{L13}. 
This fits comfortably within the upper limit of $2500
\pm^{400}_{300}\kilo\meter~\second^{-1}$ reported by \cite{L13},
which was obtained by making use of the Mach number of the NW radio relic.

\subsection{Time-since-pericenter (TSP) and the merger scenario}
\label{sec:merger_scenario}
The simulation gives two estimates for
the time-since-pericenter, with $TSP_{\rm out} = 0.46\pm^{0.09}_{0.16}~\giga \text{yr}$
and $TSP_{\rm ret} =0.91\pm^{0.39}_{0.22}~\giga\text{yr}$ for the returning model. Both the estimates of
$TSP_{\rm out}$ and $TSP_{\rm ret}$ 
fit within the observable time scale of the radio
relics, which is on the scale of $\sim1~\giga$yr.\par
\begin{figure}
	\includegraphics[width=\linewidth]{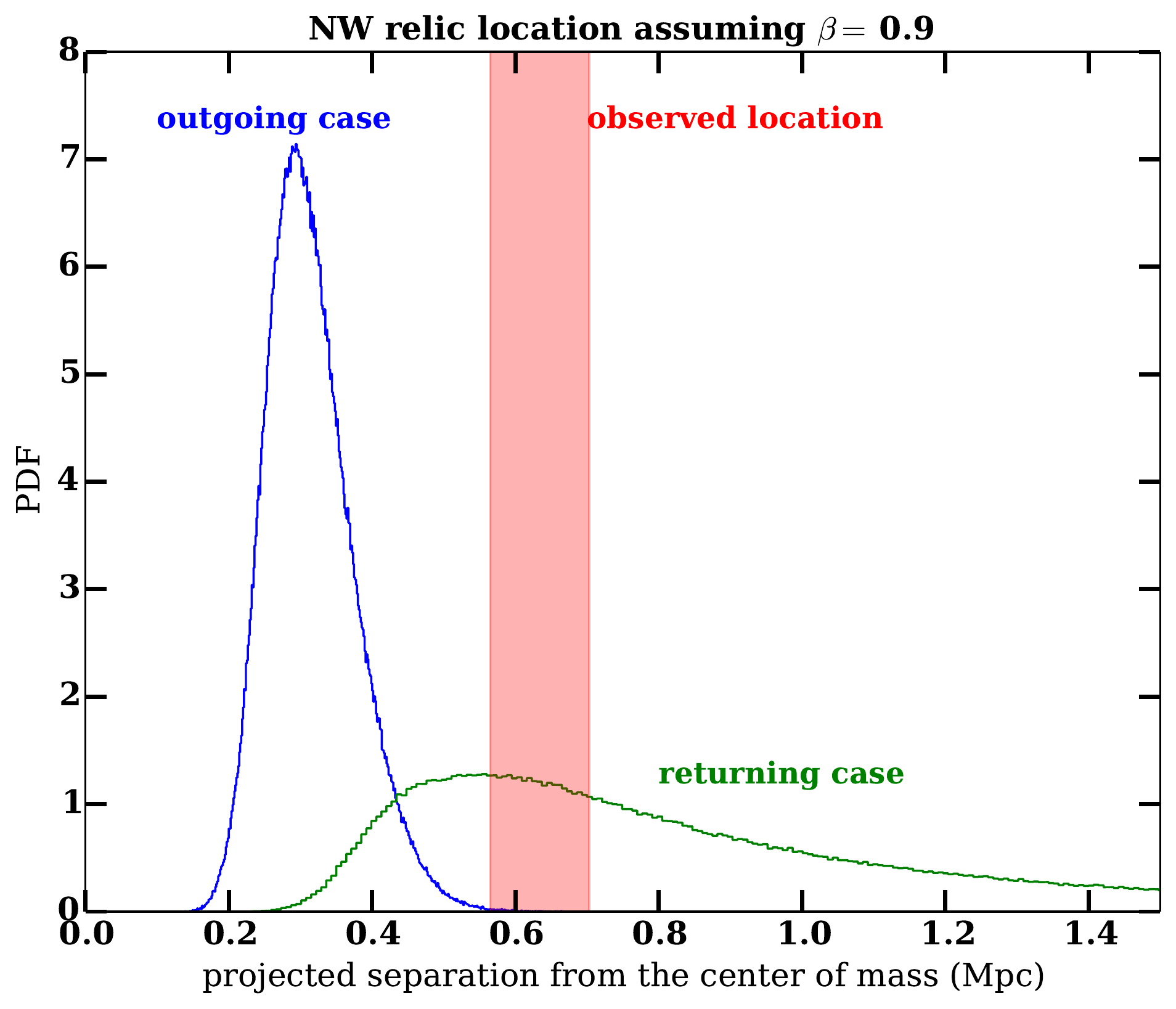}
	\caption{Comparison of the PDFs of the observed position of the NW relic (red bar
		includes the 95\% confidence interval of location of the NW radio relic in the center of mass frame) with the predicted position from the two simulated merger
		scenarios (blue for outgoing and green for the returning scenario).
	We made use of the polarization weight for producing this figure.
	The
rationale of picking $\beta = 0.9$ can be found in the last paragraph of
section 3.4.} 
	\label{fig:our_guessed_scenario}
\end{figure}

\begin{figure}
	\includegraphics[width=\linewidth]{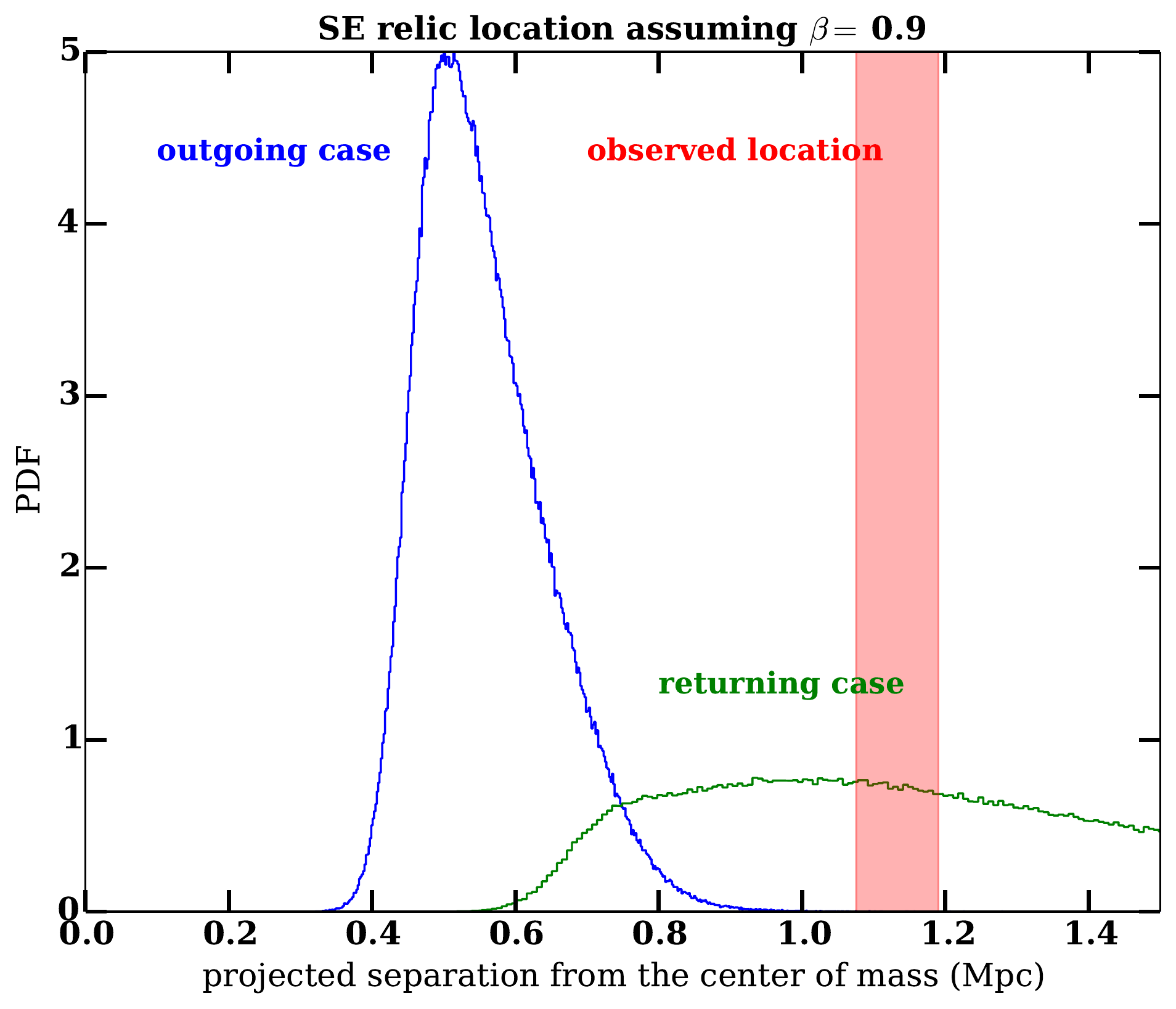}
	\caption{Comparison of the PDFs of the observed position of the SE relic (red bar
	includes the 95\% confidence interval of location of the radio relic in
the center of mass frame). We made use of the polarization weight for
producing this figure. The
rationale of picking $\beta = 0.9$ can be found in the last paragraph of
section 3.4. 
}
	\label{fig:our_guessed_scenario1}
\end{figure}

 
Based on section~\ref{sec:positionprior}, we present the PDF of
$d_{\rm proj}$ using the most likely
value of $\beta = 0.9$ in Fig.~\ref{fig:our_guessed_scenario} and
~\ref{fig:our_guessed_scenario1} to show that
the returning model is preferred for both the calculations of the NW and the
SE relic. This conclusion favoring the returning model ($M_{ret}$) holds true for the
relevant range of $  \beta
< 1.1$, which corresponds to the time-averaged velocity of the relics at
$\langle v_{NW relic} \rangle < 1000~\kilo\meter~\second^{-1}$ and $\langle
v_{SE relic} \rangle < 1800~\kilo\meter~\second^{-1}$ in the center of
mass frame. For comparison, we found that an extreme, and unlikely
range of $\beta > 1.5$ would be needed for the outgoing scenario($M_{out}$) to be
preferred. (See Appendix~\ref{app:Bayes_factor} for plots of the range
of $\beta$ that we examined). We marginalized $\beta$ to compute the
probability of the simulated relic location being compatible with the
observed location $P(S_{proj} \cap S_{obs}| M)$. We then computed   
$P(S_{proj} \cap S_{obs} | M_{ret}) / P(S_{proj} \cap S_{obs} | M_{out})$. 
The ratio of the two probabilities is found to be $\approx
2.1$ for the NW relic and a relative probability of $\approx 460$ for the
SE relic, favoring the returning scenario despite the
uncertainties. (See appendix~\ref{app:Bayes_factor}). This scenario is
further supported by the position of the cool core in the southeast as
discussed in Section 5.  
%Additionally, we
%computed the Wald statistic that tests if the sample mean (observed
%location) of the location of the relic is within the confidence interval of
%the two models, and found that the returning scenario to predict the relic
%location to be more compatible with the observed relic location. (See
%Figure~\ref{fig:waldtest}). 
%It is possible to convert the Wald statistic
%for each scenario to a p-value \citep{Wasserman04} at each $\beta$, but the
%p-value only addresses the question: ``assuming a model is true, what is 
%the conditional probability to see the relic at its observed location?''. The
%p-value itself does not constitute a comparison of the two scenarios. 
\par 
Finally, we note that the estimate of NW shock velocity at $2500
\pm^{400}_{300}~\kilo\meter~\second^{-1}$ by \cite{L13} was inferred from
the Mach number, thus, this velocity is measured in the reference frame of
the turbulent ICM, not the velocity with respect to center of
mass. Due to the difference that could arise from the different frame of
references, we have not made use of the Mach number estimate of \cite{L13} in this
calculation. If there are radio data in more frequency bands than
the radio data available now \citep{L13}, an alternative constraint of the
TSP can be constructed from the spectral aging of the electrons that were involved in the generation of the
radio relics, such as shown in \citet{Stroe14}. \par
\subsection{Sensitivity analysis of the polarization weight}
\label{sec:sensitivityTests}
We performed tests of how the output variables vary according to the
choice of the cutoff of the polarization weight between
$\alpha_{\text{cutoff}} =
29 \degree$ to $49\degree$ instead of $35 \degree$, that is, shown as the
horizontal cut off in Fig.~\ref{fig:Ensslin_fig}.
We found that in the most extreme case, choosing the cutoff values as $29
\degree$, the location of the $v_{3D}(t_{obs})$, is
increased by $16 \%$. While the $95\%$ CI of $d_{max}$ is
the most sensitive to the weight and it changes by
$\sim20 \%$ when $\alpha_{\text{cutoff}} = 49 \degree$. 
This shows that the exact choice of the cut off value for $\alpha$ for the
polarization weight does not change our estimates drastically.

\section{DISCUSSION}
%-----------------------------------------------------------------------
\subsection{Comparison of our study with other studies of El Gordo}
We outline the qualitative agreement and disagreement between our
simulations and hydrodynamical simulations of El Gordo such as
\cite{Donnert13} and \cite{Molnar14}. Our simulation focuses on giving PDF
estimates of particular dynamical and kinematic variables, whereas the
hydrodynamical simulations focused on understanding the detailed gas dynamics
required to reproduce the X-ray observables and SZ 
observables of El Gordo. The goals,
assumptions, and initial conditions of \cite{Donnert13} and \cite{Molnar14}
differ substantially with ours. However, our approach has the advantage of considering a much wider range of geometries and dynamical parameters, and is based on recently measured lensing masses.
\par 
Both hydrodynamical simulations were based on a few sets of initial
conditions, instead of thorough sampling of the inputs. For example, both
simulations made use of the mass estimates from the dynamics analysis
of \citetalias{M12} at $m_{NW} = 1.9 \times
10^{15} M_{\sun}$,
which is larger than the upper 95\% CI of the mass that we used based on
the weak lensing estimate.
Furthermore, \cite{Molnar14} initialized the relative infall velocity
(velocity when the separation of subclusters equals the sum of the two virial
radii) to be $2250~\kilo \meter~ \second^{-1}$. This corresponds to
$v_{3D}(t_{\rm per}) \gtrsim 4700~\kilo \meter~\second^{-1}$, which is close to
the escape velocity of the subclusters. 
Our simulation shows 
a negligible number of realizations with $v_{3D}(t_{\rm per}) >
3000~\kilo\meter~\second^{-1}$. 
The range of projection angles suggested by
\cite{Molnar14} of $\alpha \gtrsim 45\degree$ is also excluded by our
polarization weight, whereas we are unable to find information concerning
the projection angle of the simulation from \cite{Donnert13}.\par 
With a time resolution of 0.25 $\giga$yr,
\cite{Donnert13} gave an estimate of  $T\approx 2~\giga$yr between the
first and second core-passage in Fig. 6 of their work, while our estimate gives $T
= 1.4\pm {0.2}~\giga$yr. 
By matching the simulated X-ray luminosity and the projected separation
of $0.69$ Mpc to the corresponding observables, \cite{Donnert13} also reported their simulated work 
to best match  observations at $\sim 0.15~\giga$yr after pericenter. 
The $TSP_{\rm out}$ from \cite{Donnert13} is below the estimated 95\% CI of
$TSP_{\rm out}$ from our work.
On the other hand, \cite{Donnert13} obtained a
relative pericenter velocity between the subclusters at $\sim
2600~\kilo\meter~\second^{-1}$, which is compatible with our estimate of
$2400\pm^{400}_{200}~\kilo\meter~\second^{-1}$.  
This agreement might be due to the similar assumptions of a low
energy orbit and a small impact parameter as the initial conditions in the
work of \cite{Donnert13} and our work. 
Ideally, the hydrodynamic and Monte Carlo dynamical approaches should be combined, with new hydrodynamic simulations seeded with initial conditions motivated by the results presented here.

\subsection{Comparison to the merger scenarios of other merging clusters of galaxies}

\begin{figure}
	\includegraphics[width=\linewidth]{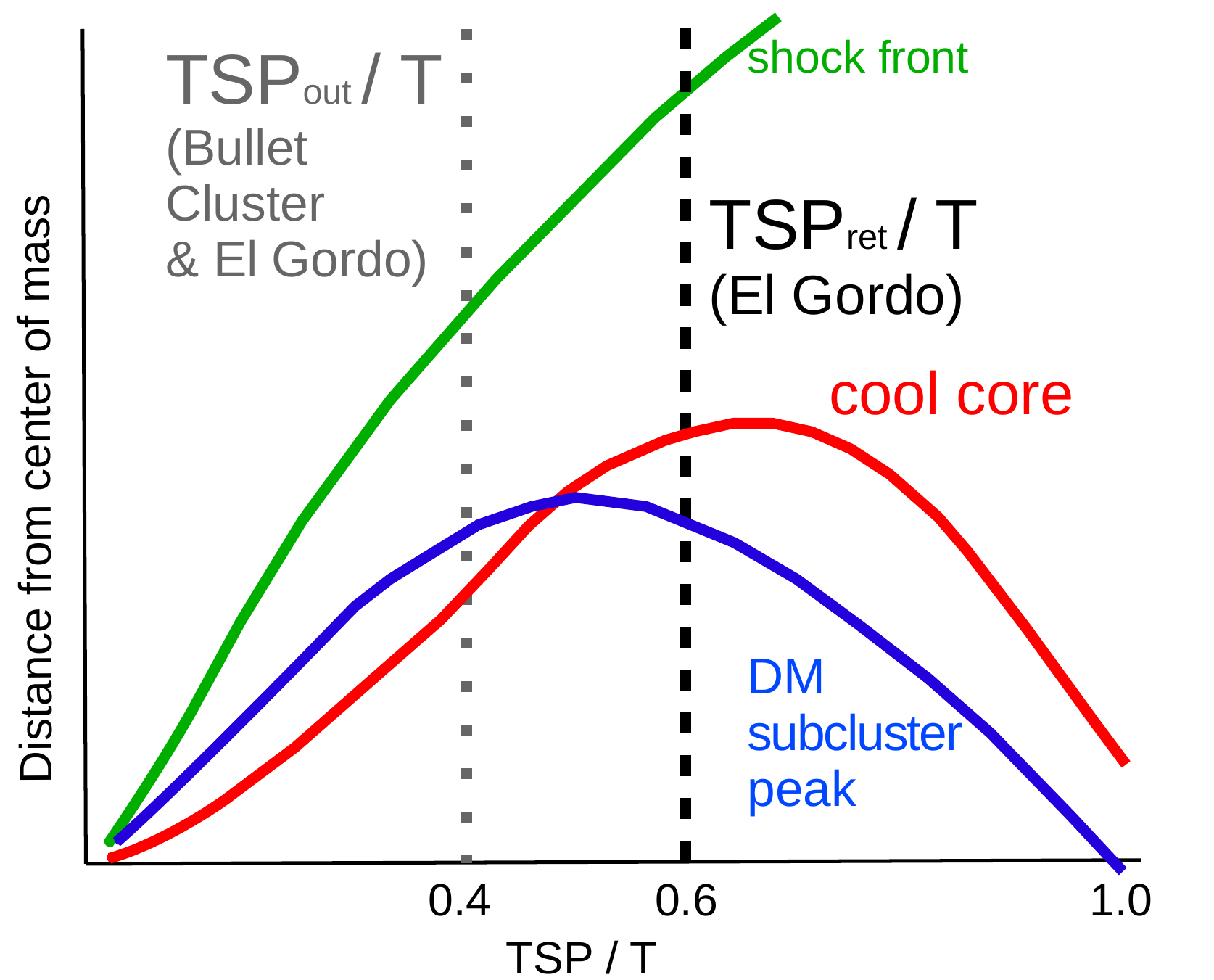}
 \caption{Schematic evolution of cool core gas and DM displacements relative to the
merger center of mass as a function of the phase ($TSP / T$), based on
simulations of a bimodal cluster merger by \citet{Mathis05}. During and
shortly after core passage, ram pressure ($=\rho v^2$) exerts substantial
force on the cool core, which then lags the DM. (This corresponds to the
outgoing scenario of $TSP_{\rm out} / T$ indicated by the grey dotted line). Ram
pressure then declines dramatically as the cool core enters regions of
lower density.  The cool core can then fall into (and past the center of)
the gravitational potential of the corresponding DM subcluster as what is
described as the slingshot effect \citep{Markevitch2007}.  The Bullet
Cluster is seen at a phase of $TSP_{\rm out} / T \approx ~0.4\text{ Gyr} / ~1.6
\text{ Gyr}$ after core passage according
to \citepalias{D13} and indeed the cool core is closest to the center of
mass. We found that El Gordo is more likely to be seen at a later stage (as
indicated by $TSP_{\rm ret} / T$ rather than $TSP_{\rm out} / T$), explaining why the DM of
El Gordo is closer to the center of mass than the cool core. \label{fig:merger_scenario}}
\end{figure}
	
	%	Illustration of the different proposed merger scenarios, i.e. projected displacement
	%	($s_{proj}$) from the center of mass of different components along the
	%	merger axis, of El
	%	Gordo and the Bullet Cluster, at different stages of the mergers.  
	%	Shortly after the core-passage
	%	phase of the merger, the cool core had to travel through a dense gas region.
	%	Ram pressure ($=\rho v^2$) would strip the gas component from a subcluster and cause the
	%	cool core to lag behind the DM peak (This corresponds to the scenario
	%	for the Bullet Cluster indicated by the grey dotted line). As time went by, the DM and the cool core of
	%	a subcluster would propagate outwards while slowing down, they would
	%	encounter a region with much lower gas density. At this later stage, 
	%	the ram pressure from the lower density gas in the
	%	outskirt region may decrease drastically, so the cool core would seem
	%	to have received a kick by a slingshot. By this
	%	later stage, the DM component would have also reached the
	%	apocenter and started returning to the center of mass for a second
	%	core-passage. The cool core would then be able to travel further away from
	%	the center of mass than the DM component (El Gordo scenario indicated
	%	by the black dashed line).
The hypothesis of El Gordo being in the returning phase is more plausible when
we compare the details of the observables of El Gordo to the Bullet
Cluster (\citealt{Bradac2006b}, \citealt{Springel2007},
\citealt{Mastropietro2008a}).
Many inferred properties are similar between the two clusters and
both clusters were observed in similar wavelengths. Both clusters are
bimodal major mergers of subclusters of substantial masses. The inferred
merger velocities are comparable at around $2600~\kilo\meter~\second^{-1}$
and $\alpha$ of both clusters are around $20 \degree$. 
In particular, the inferred outbound $TSP_{\rm out} / T \sim 0.3$ of the Bullet Cluster and El Gordo
are similar. If instead, the El Gordo
is in the returning phase of the merger (i.e. $TSP_{\rm out}$ for El Gordo is
invalid) while the Bullet Cluster is in the outgoing phase, the differences
in the observables of El Gordo and the Bullet Cluster can be explained.\par
First, the merger shock front of the Bullet Cluster is
observed only in the X-ray, but not via the radio relic, meaning that the shock may not have the
time to propagate to the outskirts of the cluster (\citealt{Bruggen2011},
\citealt{Markevitch2007}), and this bow shock is indeed observed to closely
lead the corresponding less massive subcluster by $\sim 0.08~\mega$pc,
assuming they are propagating outward. On the other hand, indirect
observables of the merger shocks of El Gordo can only be detected through the radio relic, and the shock is
further offset from the corresponding subcluster ($\sim 0.5~\mega$pc) and
the cool core ($\sim 0.4~\mega$pc). \par
%(Should discuss the Mach number,
%the speed of sounds in each cluster, and the DM velocities)  
Second, for the Bullet Cluster, the cool core (or the bullet) is closer to the
interior of the system than the corresponding less massive DM subcluster mass
peak, whereas the cool core of El Gordo is further offset from the center of
mass than the corresponding SE subcluster (See Fig.~\ref{fig:config} for
the observed positions). 
Both configurations of cool core relative to the subcluster mass peak are
mentioned in \cite{Markevitch2007}, with the case of the Bullet Cluster
explained by the ram pressure stripping effect, and the case of El Gordo
explained by the ram pressure slingshot effect, which only occurs at a
later stage of a merger (See Fig.~\ref{fig:merger_scenario} for a schema depicting this conjectured scenario).\par 
Simulation of a major merger by
\cite{Mathis05} with comparable mass ($1.4 \times 10^{15} M_{\sun}$) and
mass ratio (1:1) as El Gordo supports
our proposed scenario: it shows the turn-around of the cool core can occur after the
apocenter of the DM component, resulting in the cool core being further
away from the center of mass than the dark matter by as much as $\sim
0.2~\mega$pc.  The gas northwest of the cool core of El Gordo shows a comet-like
morphology with two tails that suggests outbound motion of the cool core, which may seem
contradictory to the returning scenario. However, from our proposed merger
scenario of El Gordo in Figure~\ref{fig:merger_scenario},
it is possible that the cool core and the DM are observed to be moving in
opposite directions, with the DM subcluster started returning for a second core-passage. If the returning scenario is true, El Gordo would be one of the first
clusters shown to be observed at a returning phase of the merger, after another
bimodal cluster merger A168 with a cool front leading the corresponding DM
subcluster \citep{Hallman04}.

\subsection{El Gordo as a probe of dark matter self-interaction}
El Gordo possesses a range of special properties that make it a promising
probe of self-interaction of DM. Its high mass ensures high DM
particle density for interactions during the high-speed core-passage. Its bimodal configuration makes it
relatively simple to interpret the offset and dynamics of the different
components. The observation of the radio relic has enabled us to
constrain the projection angle and reduce uncertainties of other dynamical
parameters. Furthermore, El Gordo is likely to be a late-stage merger
unlike other well studied clusters such as the Bullet Cluster. This gives
us a better picture of how a bimodal merger would behave at a later stage of a merger. \par 
This special merger scenario of El Gordo also raises a question: what phase
of a merger or what type of mergers would allow the most stringent
constraints on the self-interaction cross section of DM ($\sigma_{\text{SIDM}}$)? 
The use of merging clusters as probes of $\sigma_{\text{SIDM}}$ 
has been proposed and used in various papers.
(\citealt{Markevitch2004}, \citealt{Randall2008d}, \citealt{Merten2011},
\citealt{Dawson12}). One common theme among such work is
to make use of the observed offsets of the different components of the
merging clusters for the estimation. One of the most popular methods proposed by
\citealt{Markevitch2004} (method 1 in the paper) assumes the gas component would lag behind the corresponding DM
subcluster along the direction of motion due to ram pressure stripping.   
%and make use of the scattering depth $\tau_s = 1$ as an upper
%limit of the scattering depth of DM for estimating
%$\sigma_{\text{SIDM}}$. 
For El Gordo, since the cool core is further away from the
center of mass than the SE DM centroid, it is apparent that this particular
method does not apply.\par 
Alternative methods for determining the self-interaction cross sections,
such as from the galaxy-DM offset, are yet to be perfected. Future work is
required to investigate how to best characterize the spatial distribution
of the galaxies. One pending question is to investigate if the luminosity
density peak or number density peaks would better represent the galaxy
distributions. The galaxy number density map of El Gordo
\citepalias{Jee13} shows a noteworthy $\sim0.2~\mega$pc offset between the SE
galaxy number density peak and the SE DM centroid, while there is almost no offset
between the corresponding luminosity peak and DM centroid. The discrepancy
between the number density peak and the luminosity peak is due to a very
bright brightest-cluster-galaxy (BCG) located close to the corresponding DM peak.
The BCGs tend to mark the bottom of the potential, so this further supports the
ram-pressure slingshot scenario outlined here. At the same time, this
illustrates the need for further understanding of the behavior of the
galaxy number and luminosity densities in dynamic situations before
galaxy-DM offsets can be used to infer DM properties. 
\par 
\subsection{Improving constraints of merger scenario using prior knowledge from simulations}
This work has allowed us to examine
what information would be needed to better understand the merger
dynamics and scenario. Before this work, simulations of merger shocks have
focused on providing estimates of the local conditions of the physics
responsible for the generation of the radio relic or the gas physics. In this work, we demonstrated that the global
properties of the shocks, are also important for understanding the merger scenario. 
Important questions concerning merging galaxy
clusters pending for answers include:  
\begin{itemize}
\item What are the typical propagation velocities of the shock wave that
	corresponds to the radio relic {\it in the center of mass (CM) frame} of the cluster?
\item What physical properties of the DM subclusters would correlate the
	best with the time-evolution of the propagation velocity of the shock
	wave (in the CM frame)?  
\item What is the typical duration
after the merger for which radio relics are observable in terms of the merger
core-passage time-scales? 
\item How generalizable is the merger scenario in Figure
~\ref{fig:merger_scenario}?  
\item How would the galaxy-DM offset evolve if we were to add that information
	to Fig.~\ref{fig:merger_scenario}?
\item For how long do transcient X-ray features in merging clusters (such
		as the wake in El Gordo) persist?
\end{itemize}
We urge simulators to narrow the gap between simulations and data by
investigating these issues.
\section{SUMMARY} 
We provide estimates of the dynamical parameters of El Gordo using Dawson's
Monte Carlo simulation, in particular, we 
\begin{enumerate}
	\item demonstrated the first use of polarization fraction information from
		the radio relics to reduce our estimates of the projection angle from
		$43\degree \pm ^{26}_{24}$ to $21 \degree \pm^{9}_{11}$ (See
		Fig.~\ref{fig:geom_geom}). By performing sensitivity analysis, we
		showed that this weighting function helps reduce uncertainty for the dynamical
		variables without changing the dynamical variable estimates drastically ($< 20\%$).\\ 
	\item inferred the {\it relative} pericenter velocity 
		between the subclusters of El Gordo as 
		$2400\pm^{400}_{200}~\kilo\meter~\second^{-1}$ \\ 
	\item showed that a returning scenario is favored if $\langle v_{NW relic}\rangle \leq
		1000~\kilo
		\meter~\second^{-1}$ and $\langle v_{SE relic}\rangle \leq
		1800~\kilo\meter~\second^{-1}$
where the velocities are in the CM frame and angle brackets denote averaging over the time since pericenter
		It takes
		an unlikely high speed of $\langle v_{relic} \rangle \gg 1.5~v_{3D,
		sub}(t_{\rm per})$ for the outgoing scenario to be favored. \\ 
	\item showed how our inferred
		returning scenario may explain the unexpected location of the cool
		core, namely, the cool core being close to the center of mass of the
		cluster, and still be consistent with the wake / gas-tail morphology of the cool core. 
\end{enumerate}
As large scale sky surveys come online, more cluster mergers at late
stages of their merger will be discovered. El Gordo will serve as one of
the best studied examples of a bimodal cluster merger for comparison.  
\section{ACKNOWLEDGEMENTS}
We thank Franco Vazza, Marcus Br\"{u}ggen and Surajit Paul for sharing
their knowledge on the simulated properties of radio relic and merger
shocks. We extend our gratitude to Reinout Van Weeren for first proposing the use of
radio relic to weight the Monte Carlo realizations. We appreciate the
comments from Maru\v{s}a Brada\v{c} about using the position of the relic to
break degeneracy of the merger scenario. KN is grateful to Paul Baines and
Tom Loredo for discussion of the use of prior information and sensitivity tests. 
Part of this work was performed under the auspices of the U.S. DOE by LLNL
under Contract DE-AC52-07NA27344. 
JPH gratefully acknowledges support from Chandra (grant number GO2-13156X)
and Hubble (grant number HST-GO-12755.01-A).
We would also like to thank 
GitHub for providing free repository for version control for our data and
analyses. This research made use of APLpy, an open-source plotting package for Python
hosted at http://aplpy.github.com; Astropy, a community-developed core
Python package for Astronomy \citep{astropy}; AstroML, a
machine learning library for astrophysics \citep{VanderPlas2012}, and IPython, a system for
interactive scientific computing, computing in science and engineering
\citep{Perez2007}.\par
Note: The authors have made the Python code for most of the analyses openly
available at
\href{https://github.com/karenyyng/ElGordo\_paper1}{https://github.com/karenyyng/ElGordo\_paper1}. 

\bibliographystyle{mn2e}
\bibliography{ElGordo1}
\appendix
\section{DEFAULT WEIGHTS USED FOR DAWSON'S MONTE CARLO SIMULATION}
\label{app:priors}
The default weight that we employed can be summarized as
follows for most of the output variables: 
\begin{equation}
	w(v_{3D}(t_{\rm per})) = 0\text{ if }v_{3D}(t_{\rm per}) >
	v_{\text{free fall}}. 
\end{equation}
\begin{equation}
	w(TSP_{\rm out}) = 
	\begin{cases}
		& \text{const}~\text{if }TSP_{\rm out} < \text{age of universe at } z=0.87	\\
		& 0~\text{otherwise}.
	\end{cases}
\end{equation}
In addition, we apply the following weight on $TSP_{\rm ret}$:
\begin{equation}
	w(TSP_{\rm ret}) = 
	\begin{cases}
		& \text{const}~\text{if }TSP_{\rm ret} < \text{age of universe at } z=0.87	\\
		& 0~\text{otherwise} \label{eqn:TSM_1}.
	\end{cases}
\end{equation}
To correct for observational limitations, we further convolved the
posterior probabilities of the different realizations with 
\begin{equation}
	w(TSP_{\rm out} | T) = 2 \frac{TSP_{\rm out}}{T},
\end{equation}
to account for how the subclusters move faster at lower $TSP$ and thus it
is more probable to observe the subclusters at a stage with a larger $TSP$.
\par 
\section{PLOTS OF OUTPUTS OF THE MONTE CARLO SIMULATION}
\begin{figure}
	\begin{center}
	\includegraphics[width=\linewidth]{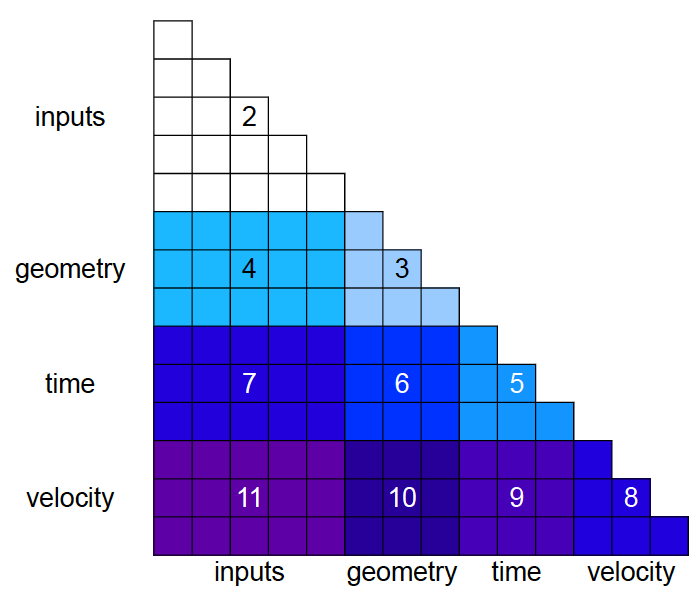}
	\end{center}
	\caption{Matrix of variables used in the simulations. We present them in
	4 categories, including, inputs, geometry, time and velocity. Regions of
	the same color represent one plot and the number
indicates the corresponding figure number in this appendix.
\label{fig:plot_config}
}
\end{figure}

We present the PDFs of the inputs of the dynamical simulation and the
marginalized PDFs of the outputs after applying the polarization weight in
addition to the default weights. See Fig.~\ref{fig:plot_config} for explanations of
the order that we present the figures containing the PDFs . 
\label{app:results}
%%%%%%%%%%%%% TASK --- 
\clearpage
\begin{figure*}
	\begin{minipage}{180mm}
	\begin{center}
	\includegraphics[width=0.65\linewidth]{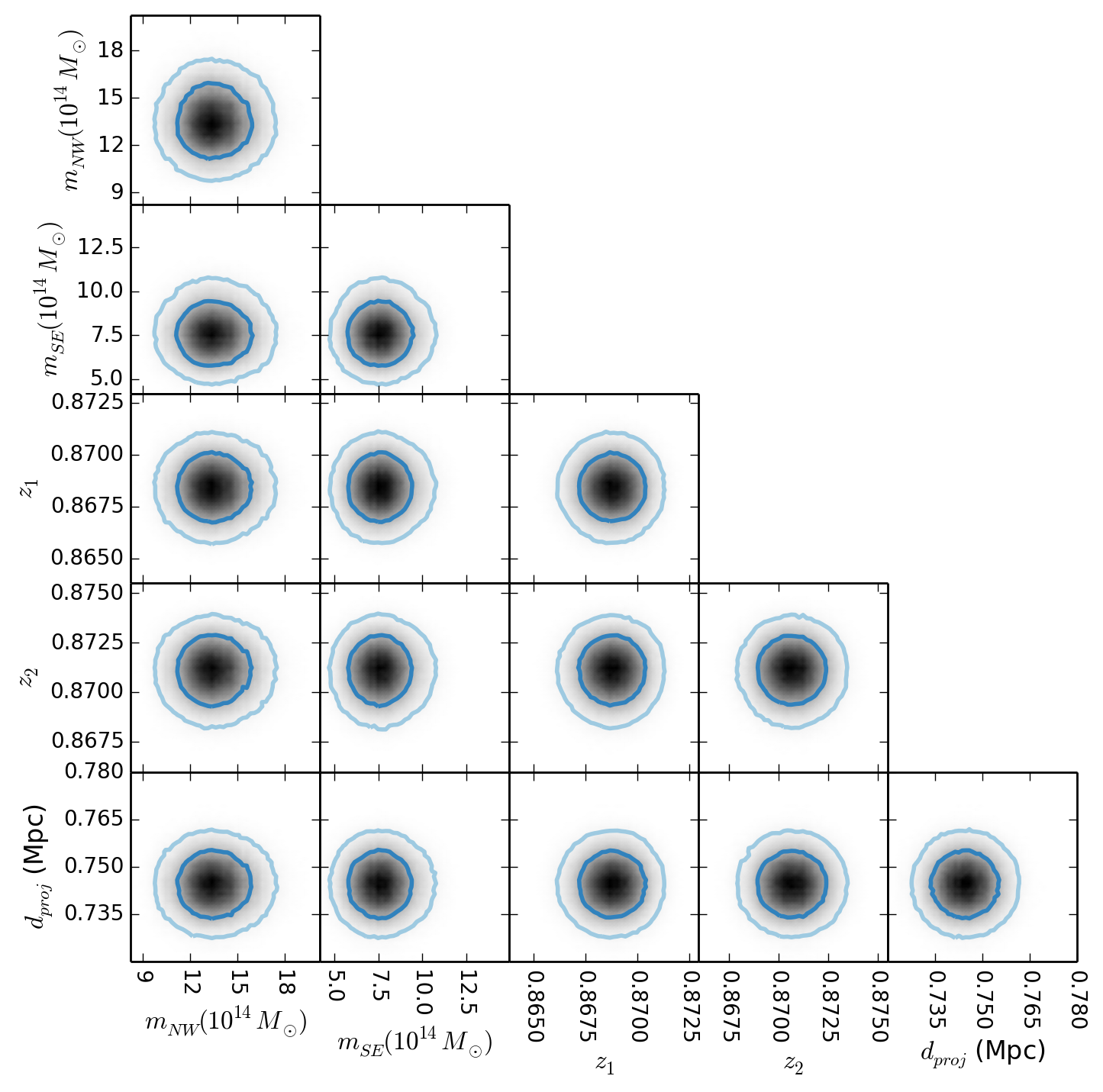}
	\caption{Marginalized 2-dimensional PDFs of original inputs (vertical axis) 
		and the inputs after applying polarization weight and default weights 
		(horizontal axis). The inner and outer contour
denote the central 68\% and 95\% confidence regions respectively.
The circular contours show that the application of weights did not introduce
uneven sampling of inputs. }
	\end{center}
	\end{minipage}
\end{figure*}
\begin{figure*}
\begin{minipage}{180mm}
	\begin{center}
	\includegraphics[width=0.5\linewidth]{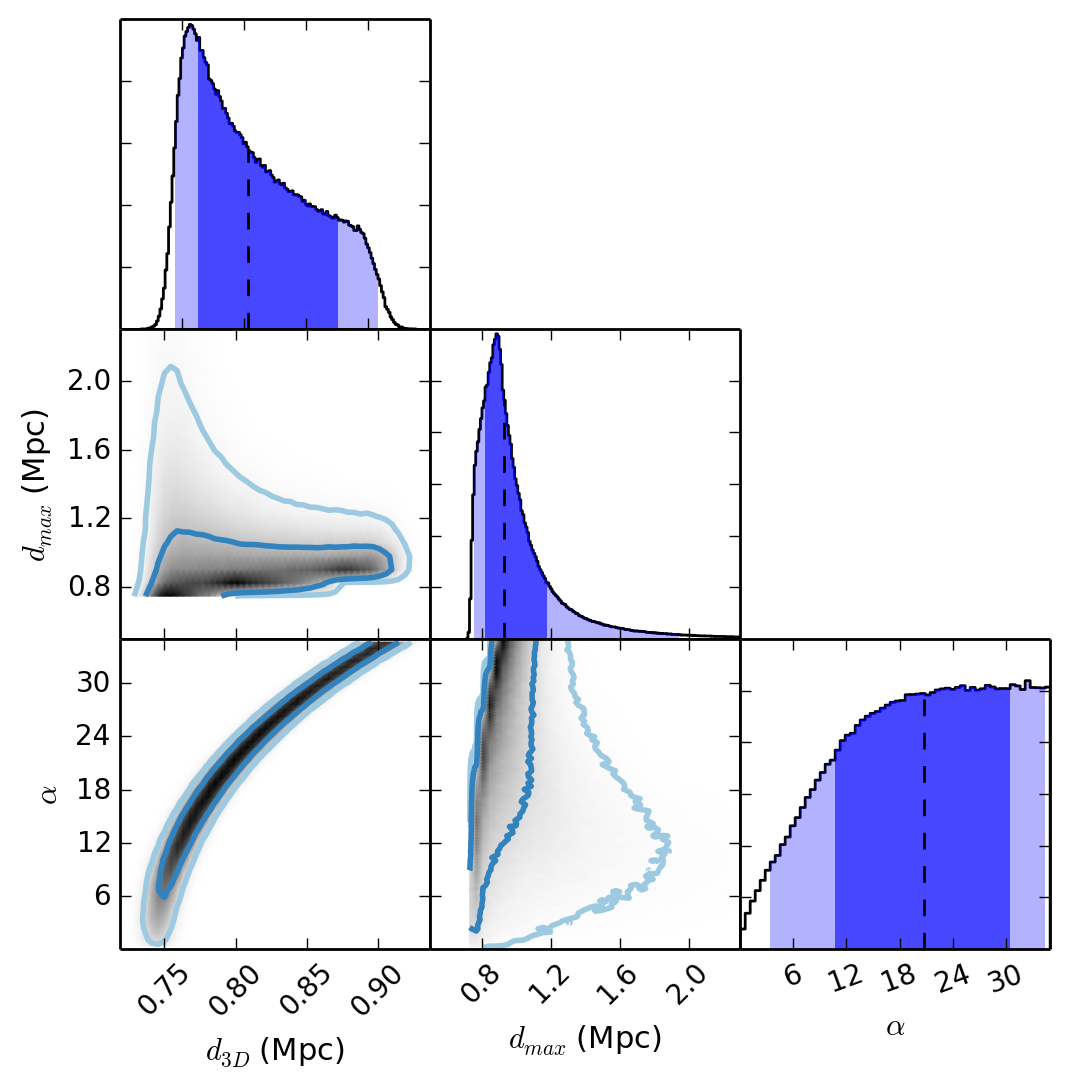}
	\caption{One-dimensional marginalized PDFs (panels on the diagonal) and
		two-dimensional marginalized PDFs of variables
		denoting characteristic distances and projection angle of the mergers.
	\label{fig:geom_geom}
	}
	\end{center}
	\end{minipage}
\end{figure*}
\begin{figure*}
\begin{minipage}{180mm}
	\begin{center}
	\includegraphics[width=0.7\linewidth]{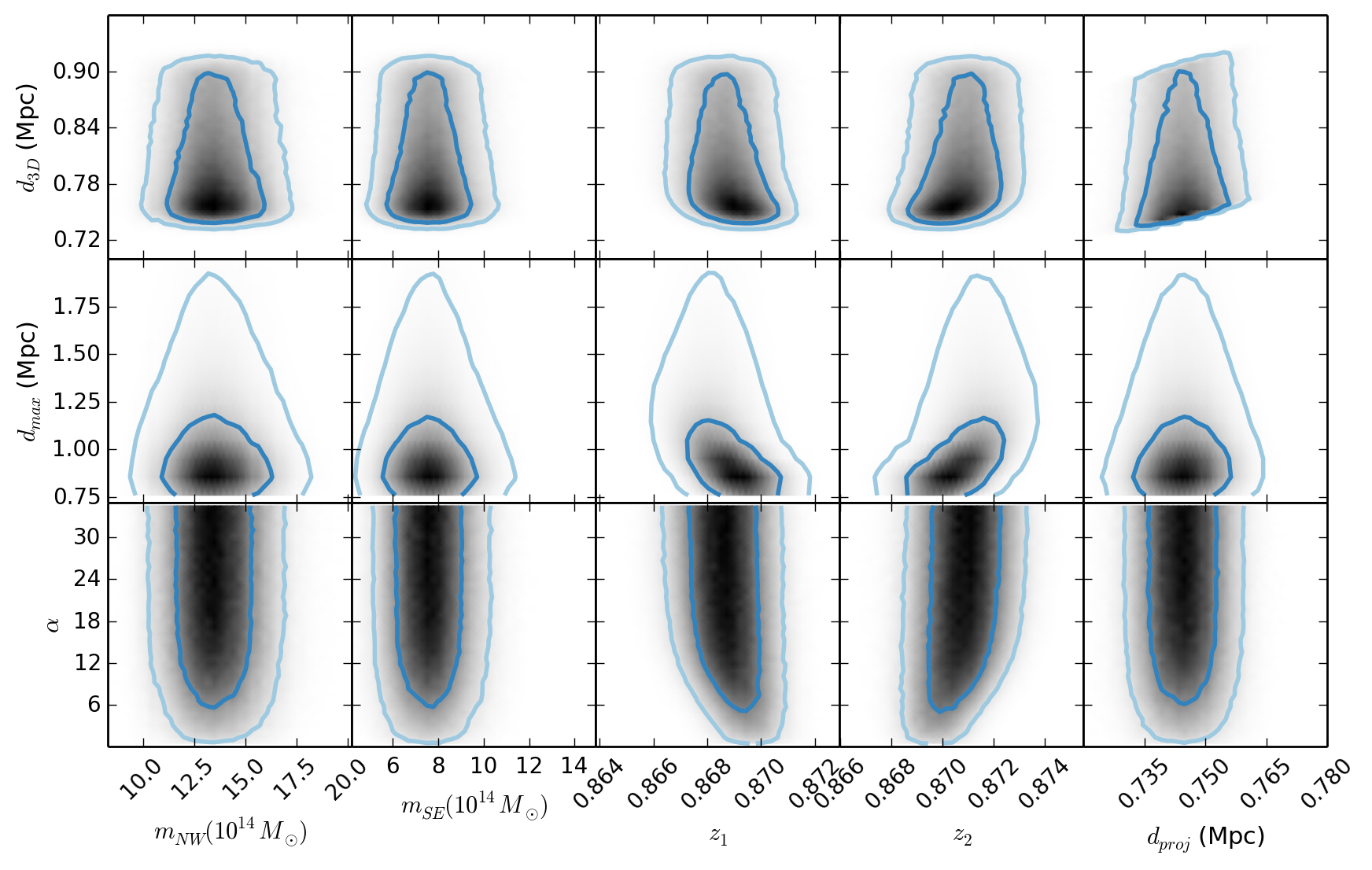}
	\caption{Marginalized PDFs of characteristic distances and projection
		angle of the merger and the inputs of the simulation.}
	\end{center}
	\end{minipage}
\end{figure*}
\begin{figure*}
\begin{minipage}{180mm}
	\begin{center}
	\includegraphics[width=0.5\linewidth]{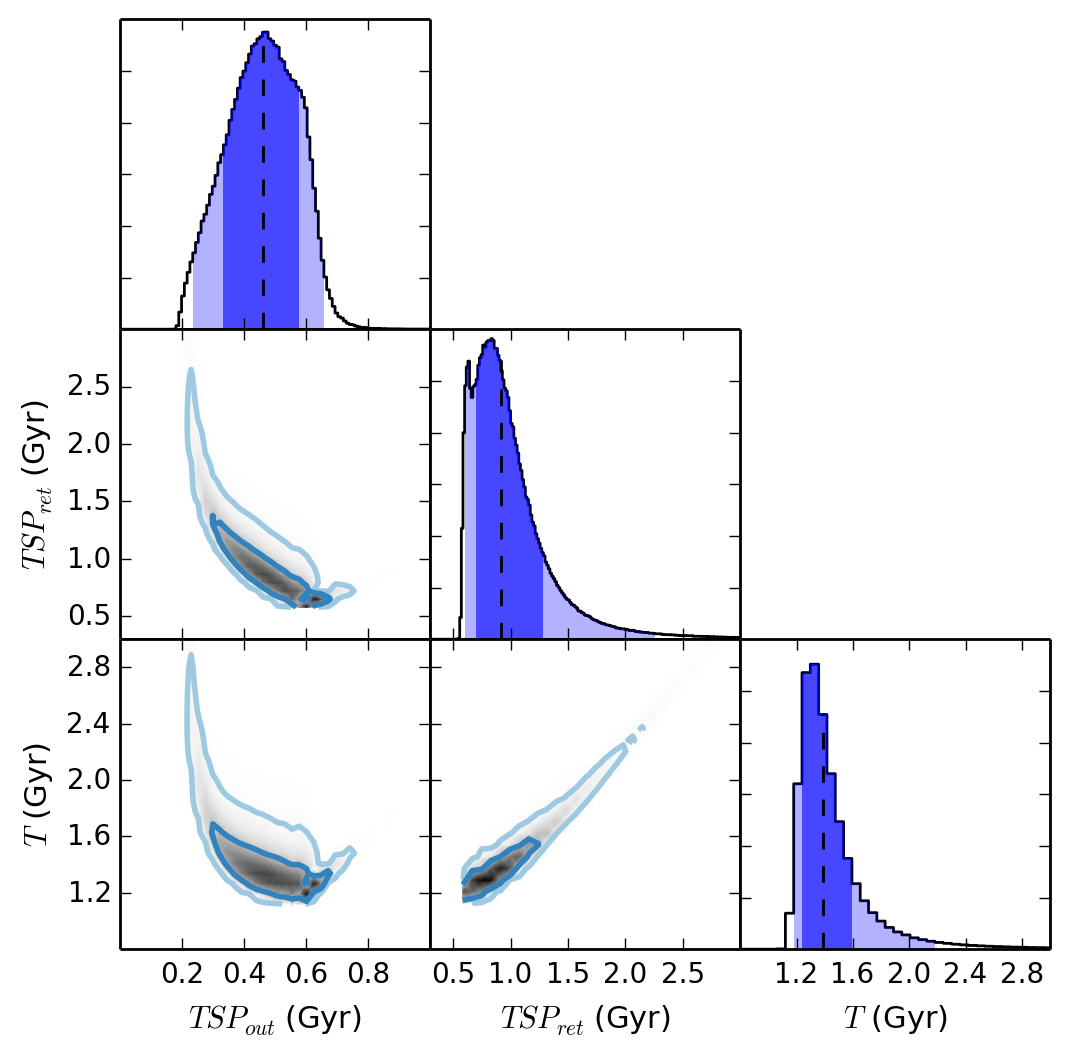}
	\caption{One-dimensional PDFs of characteristic timescales of the simulation
(panels on the diagonal) and the marginalized PDFs of different
timescales. Note that there is a default weight for constraining $TSP_{\rm out}$ but
not for $TSP_{\rm ret}$ and $T$, so $TSP_{\rm out}$ spans a smaller range.}
	\end{center}
\end{minipage}
\end{figure*}
\begin{figure*}
\begin{minipage}{180mm}
	\begin{center}
	\includegraphics[width=0.5\linewidth]{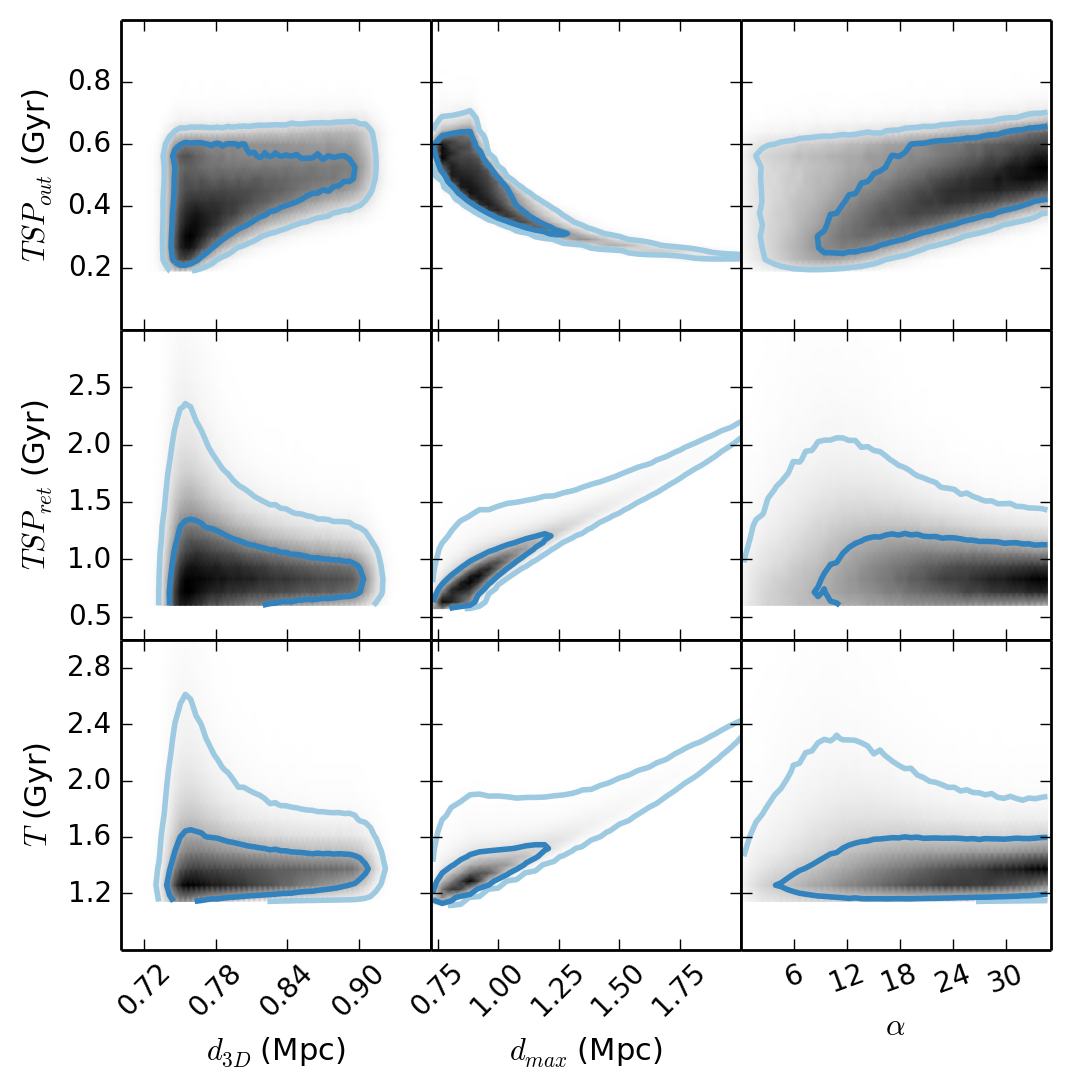}
	\caption{Marginalized PDFs of characteristic timescales of the simulation
and the characteristic distances and the projection angle of the merger. }
	\end{center}
\end{minipage}
\end{figure*}
\begin{figure*}
\begin{minipage}{180mm}
	\begin{center}
	\includegraphics[width=0.7\linewidth]{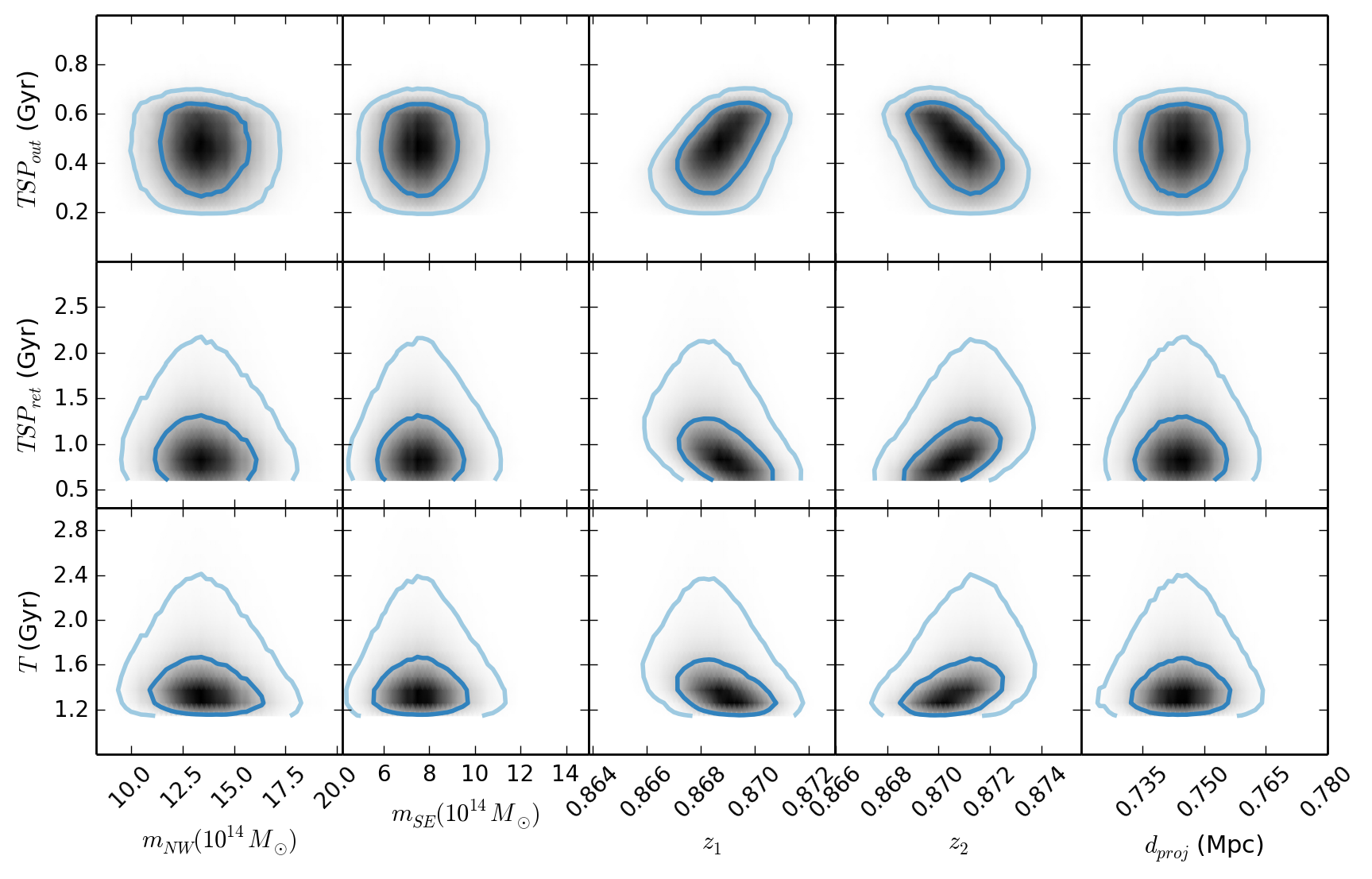}
	\caption{Marginalized PDFs of characteristic timescales of the simulation
and the inputs.}
	\end{center}
\end{minipage}
\end{figure*}
\begin{figure*}
\begin{minipage}{180mm}
	\begin{center}
	\includegraphics[width=0.5\linewidth]{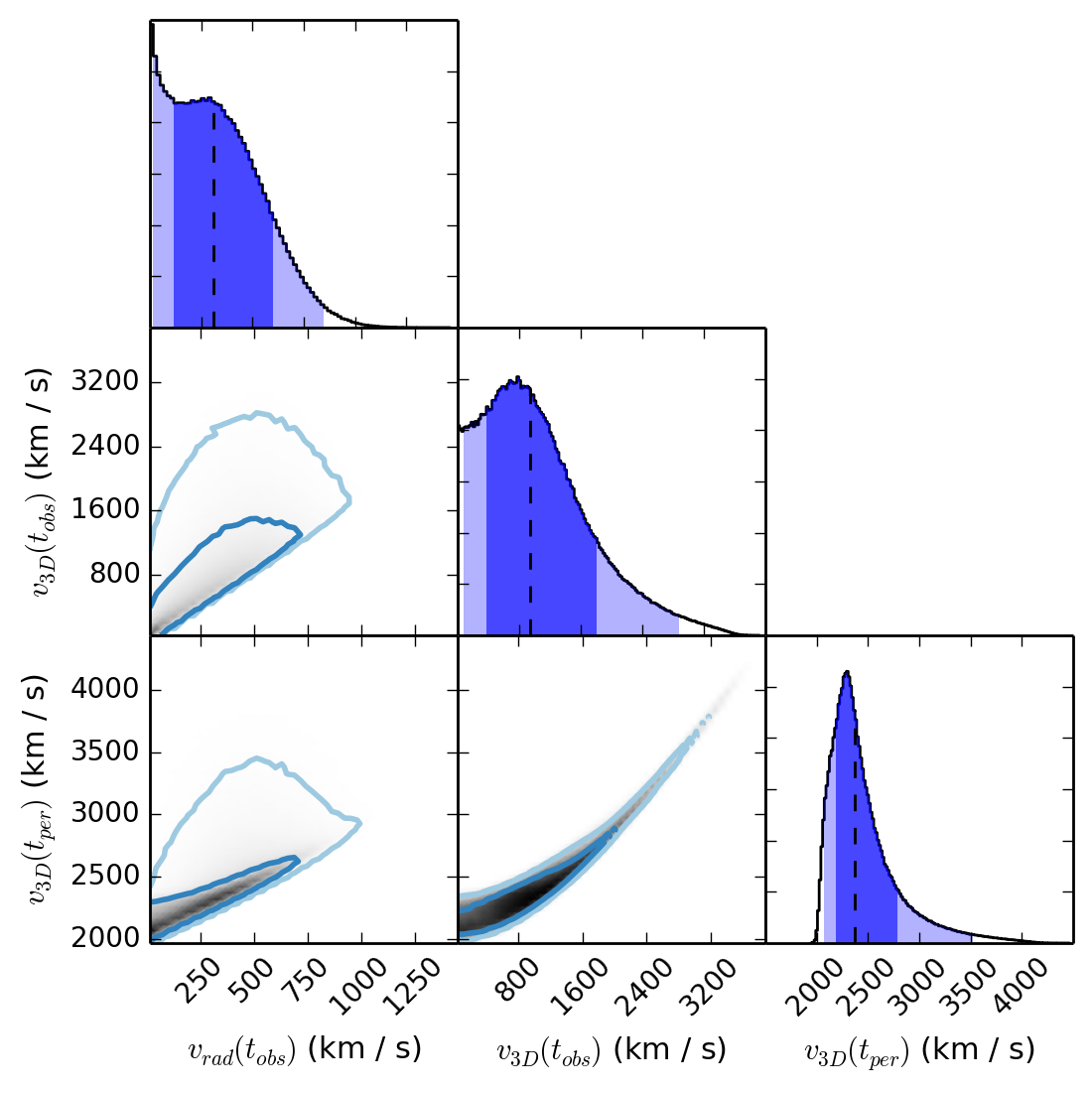}
	\caption{One-dimensional marginalized PDFs of velocities at
	characteristic times (panels on the diagonal) and marginalized PDFs of
different velocities.}
	\end{center}
\end{minipage}
\end{figure*}
\begin{figure*}
\begin{minipage}{180mm}
	\begin{center}
	\includegraphics[width=0.5\linewidth]{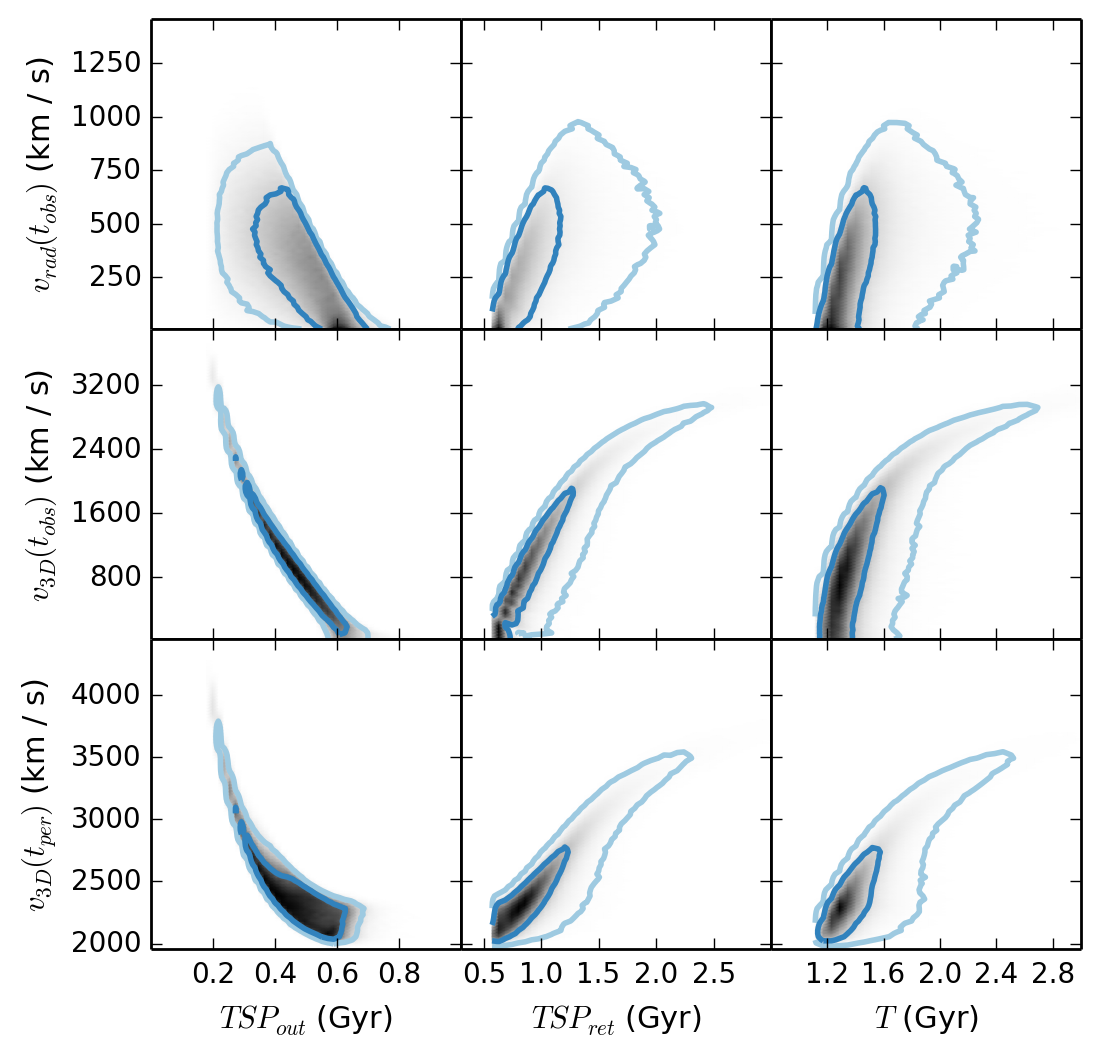}
	\caption{Marginalized PDFs velocities and the characteristic timescales
	of the simulation against the inputs.}
	\end{center}
\end{minipage}
\end{figure*}
\begin{figure*}
\begin{minipage}{180mm}
	\begin{center}
	\includegraphics[width=0.5\linewidth]{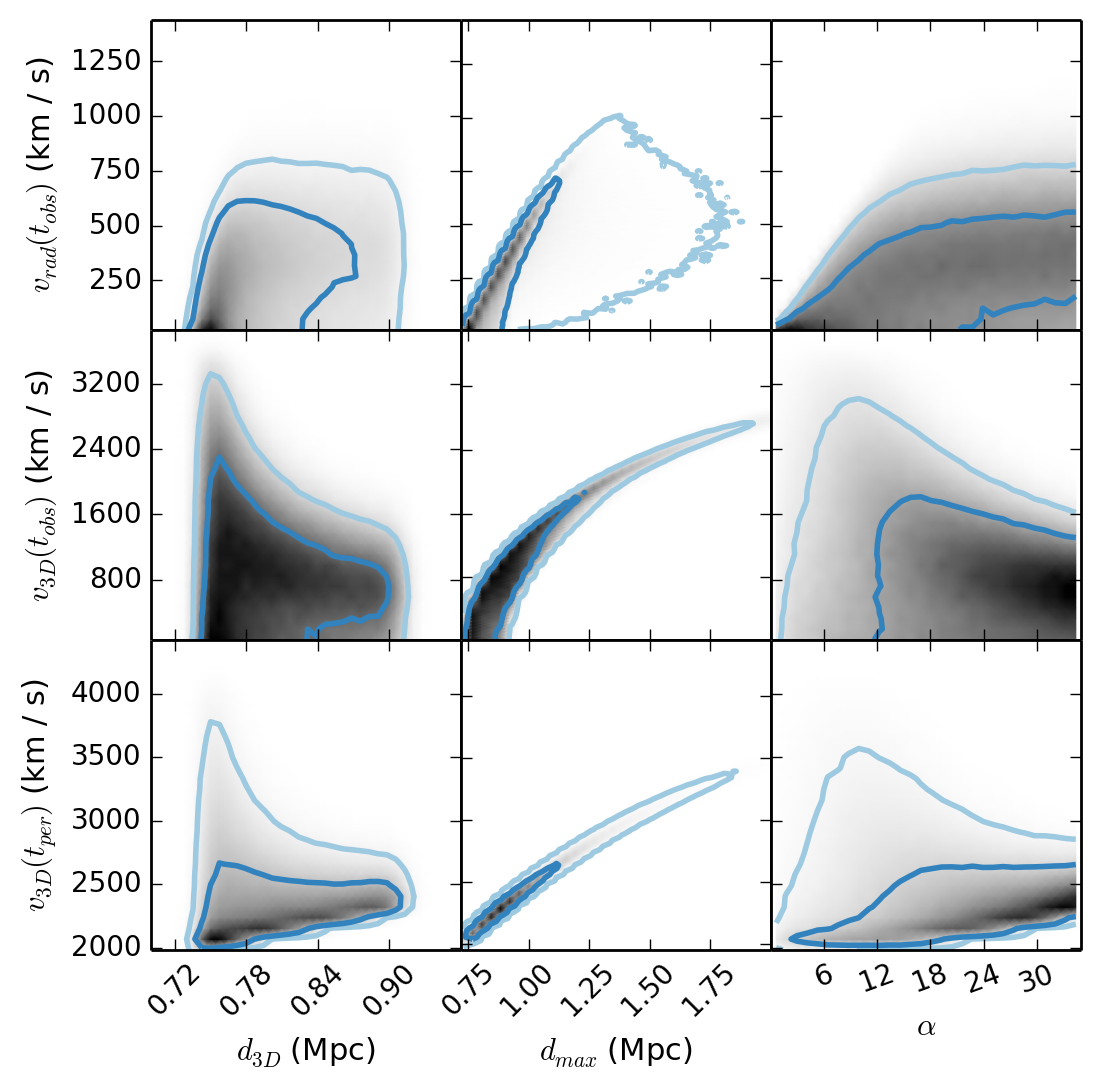}
	\caption{Marginalized PDFs of the velocities at characteristic timescales
		and the characteristic distances and the projection angle of the merger. }
	\end{center}
\end{minipage}
\end{figure*}
\begin{figure*}
\begin{minipage}{180mm}
	\begin{center}
	\includegraphics[width=0.7\linewidth]{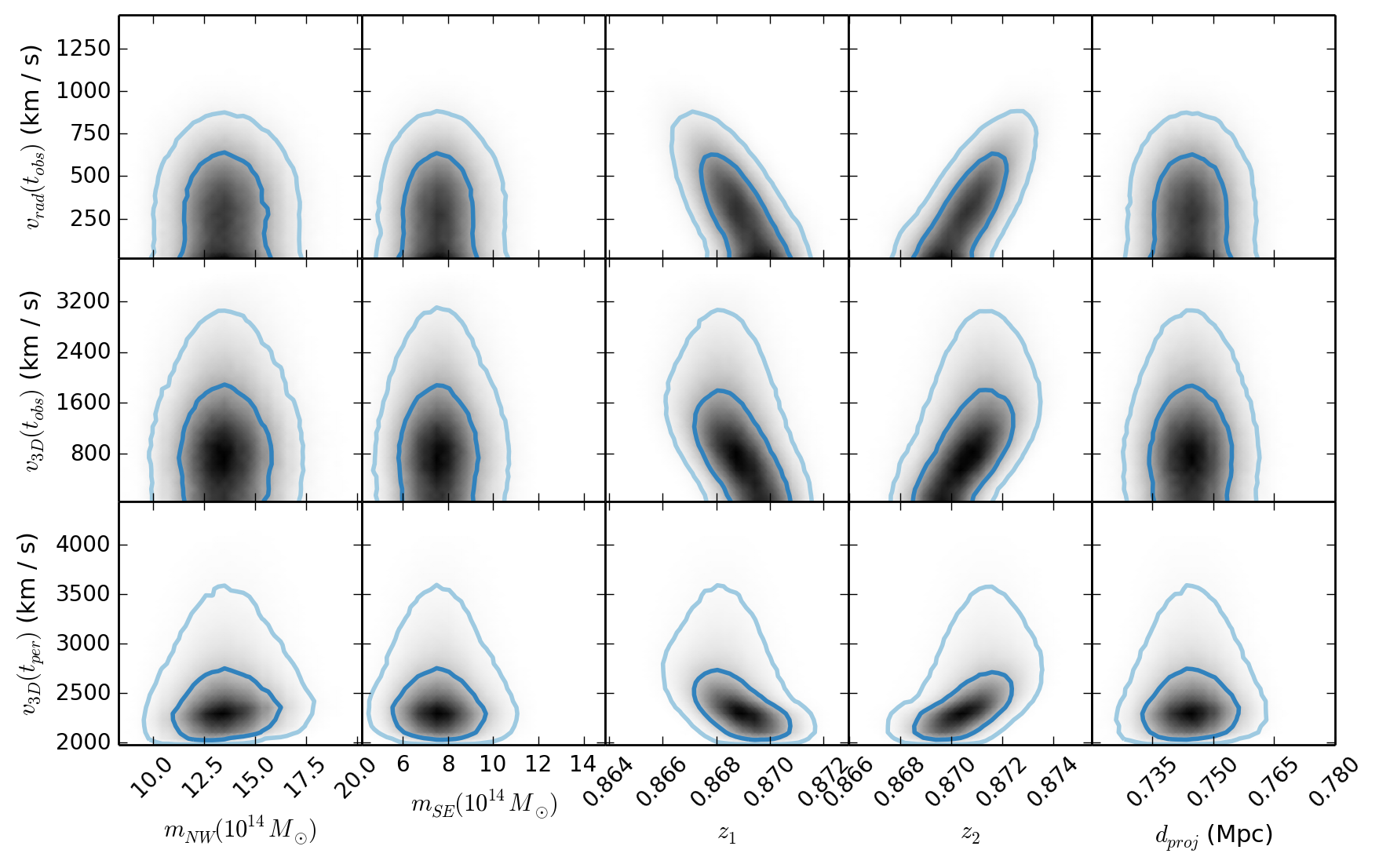}
	\caption{Marginalized PDFs of relative velocities characteristic
	timescales of the simulation and the inputs.}
	\end{center}
\end{minipage}
\end{figure*}

\section{COMPARISON OF THE OUTGOING AND RETURNING SCENARIO}
\label{app:Bayes_factor}
Here, we compare the different merger scenarios using the two relics
independently and show that they consistently give the conclusion that the returning
scenario is favored for the plausible range of $\beta$. For each merger
scenario, we compute
the (marginalized) probability of producing simulated
values ($s_{proj}$) compatible with the observed location of the radio
relic ($s_{obs}$). 

Quantitatively, we want to compute and compare the probability:  
\begin{align} 
	&P(s_{proj} \text{ compatible with }s_{obs} | M)  \label{eqn:prob}\\
	&=\iint f(S_{proj} \cap S_{obs} | M, \beta) f(\beta | M) d s_{proj} d\beta\\
	&=\iint  f(s_{proj}|M, \beta) f(s_{obs}) f(\beta | M) d s_{proj}
	d\beta, 
\end{align}
where $f$ indicates the corresponding PDF, $M$
represents one of the merger scenarios, and $\beta$ is defined in equation
~\ref{eqn:NW_speed}, $s_{proj} \in S_{proj}$ and $s_{obs} \in S_{obs}$. 
We set our priors set to be uniform for the marginalization: 
\begin{equation}
	f(\beta | M_{ret}) = f(\beta | M_{out}) =  
	\begin{cases}
		& \text{const}~\text{if } 0.7 \leq \beta \leq 1.5 \\
		& 0~\text{otherwise}.
	\end{cases}
\end{equation}
which is more conservative than the most likely range of $\beta$, which is
$0.7 < \beta < 1.1$.
We found: 
\begin{align}
	&P(S_{proj} \cap S_{obs} | M_{ret}) / P(S_{proj} \cap S_{obs} |
	M_{out})\\
	&=
 \begin{cases}
  2.1 \text{ for the NW relic},\\
  460 \text{ for the SE relic},
 \end{cases}
\end{align}
which shows that the returning scenario is favored over the outgoing
scenario. 

This test quantity differs from the traditional hypothesis testing or model comparison in several ways: 
\begin{enumerate}
\item we did not compute a likelihood function.  We have adopted
non-parametric PDFs in our Monte Carlo simulation,i.e. there is no well-known
functional form of the likelihood in our context. We make use of
$f(S_{proj} \cap S_{obs})$ to penalize the simulated values being different from our observed data
\item with this quantity, we are not asking whether the expected value of
	the radio relic such as the mean or median from each model match the
	observation best. Those estimators take into account the values that do not match the observed location of the radio relic. 
\item we marginalized the uncertainty in $\beta$ to be as conservative
	as possible, instead of assuming a fixed value of $\beta$.
\end{enumerate} \par
%The second quantity that we compute is the Wald statistic at a range of possible $\beta$ values.
%Wald statistic is usually used hypothesis testing.
%Computing the Wald statisic allows us to ask, given a particular model (merger
%scenario),
%whether the null value (sample mean) is in the confidence
%interval \citep{Wasserman04}. The Wald statistic that we compute for each model is: 
%\begin{align}
%	W(\beta) = \frac{\bar{s}_{obs} - \mu(\beta)}{\sigma / (n)^{1/2}} 
%\end{align}
%where $\bar{x}$ is the sample mean, $\mu(\beta)$, $\sigma$ and $n$ are the
%population mean, population standard deviation and size of samples of the
%simulated model respectively. A higher Wald statistic value would represent
%a larger difference between the observed and model value, while taking into account the model uncertainty. Within the range of
%most likely $0.7 < \beta < 1.5$, the Wald statistic shows that the observed
%relic location is more compatible with the confidence interval of the returning
%scenario. (See Figure~\ref{fig:waldtest})

\begin{figure}
	\includegraphics[width=\linewidth]{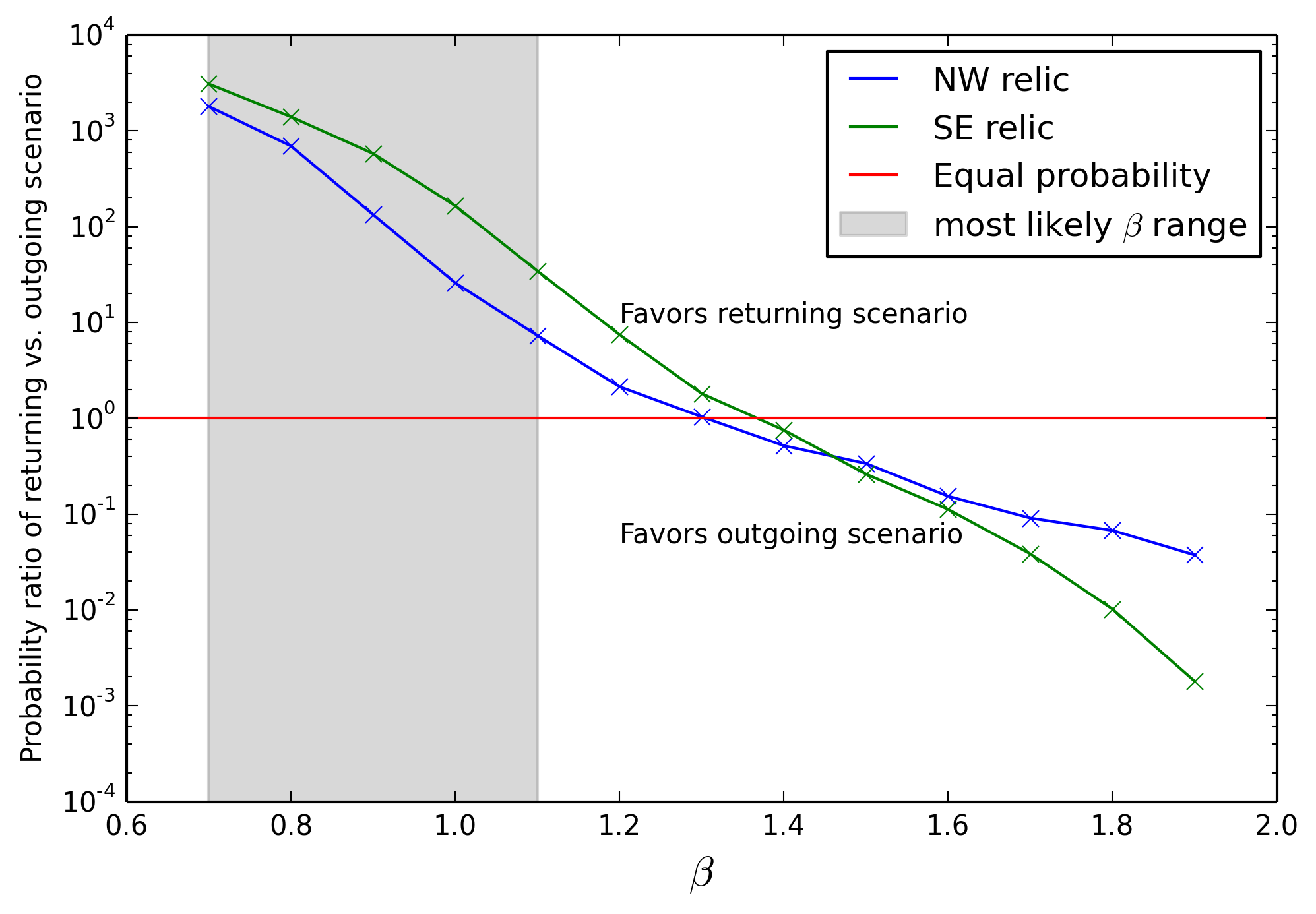}
	\caption{Probability ratio between the returning model (numerator) and
		the outgoing model at given $\beta$. We remind readers $\beta$ is a factor
		relating the {\it time-averaged} shock velocity and the pericenter
		velocity of the corresponding subcluster.  
	\label{fig:prob_ratio}}
\end{figure}

\begin{figure}
\includegraphics[width=\linewidth]{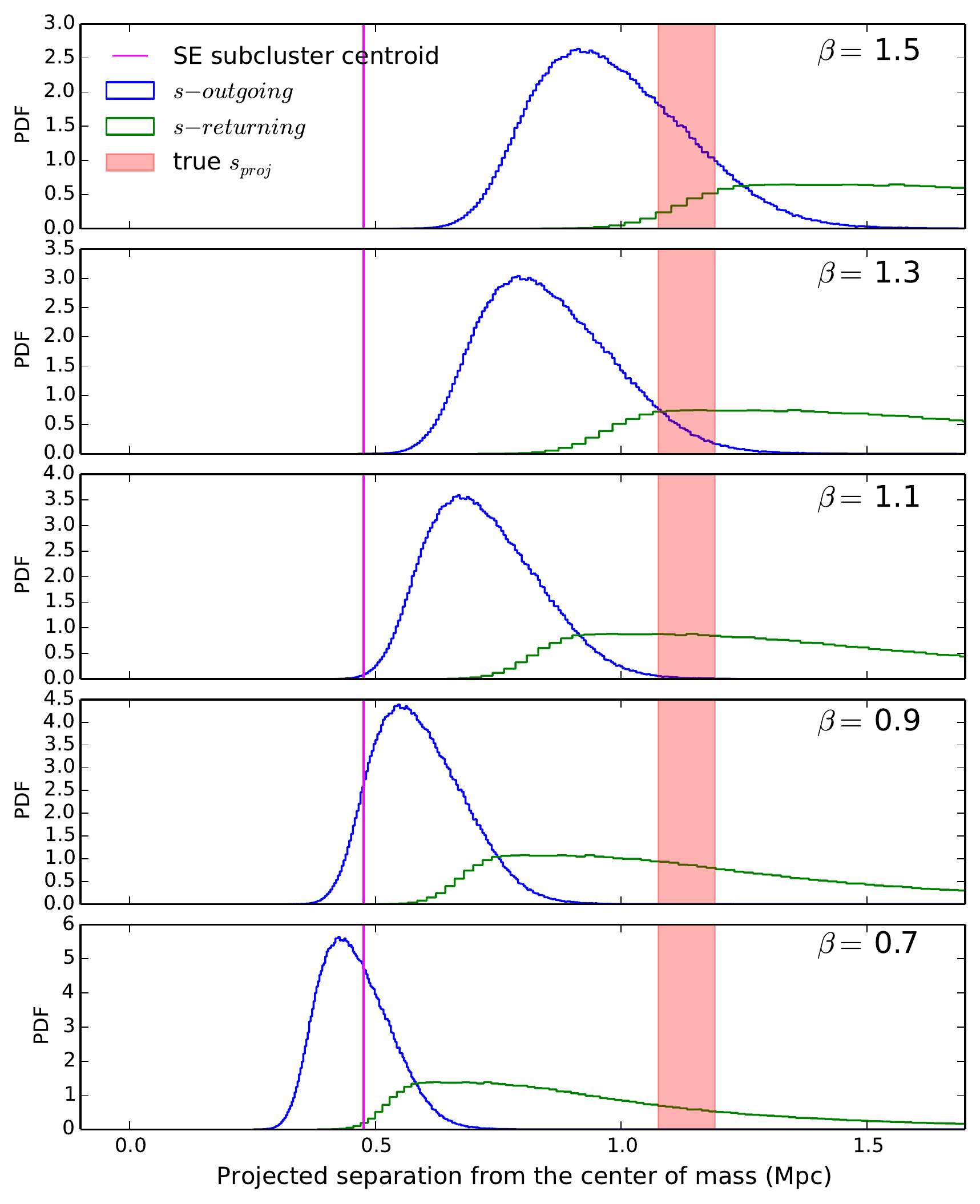}
	\caption{Comparison of the PDFs of the observed position of the SE relic (red bar
		includes 95\% confidence interval of location of relic in the center of
		mass frame) with the predicted position from the two simulated merger scenarios (blue for outgoing and green for the returning scenario). 
	For the plausible values of $\beta < 1.1$, the returning scenario is preferred. 
	We obtained similar conclusion about the merger scenario as for the NW
	relic calculation.
	\label{fig:positionprior_SE}}
\end{figure}

\begin{figure}
	\includegraphics[width=\linewidth]{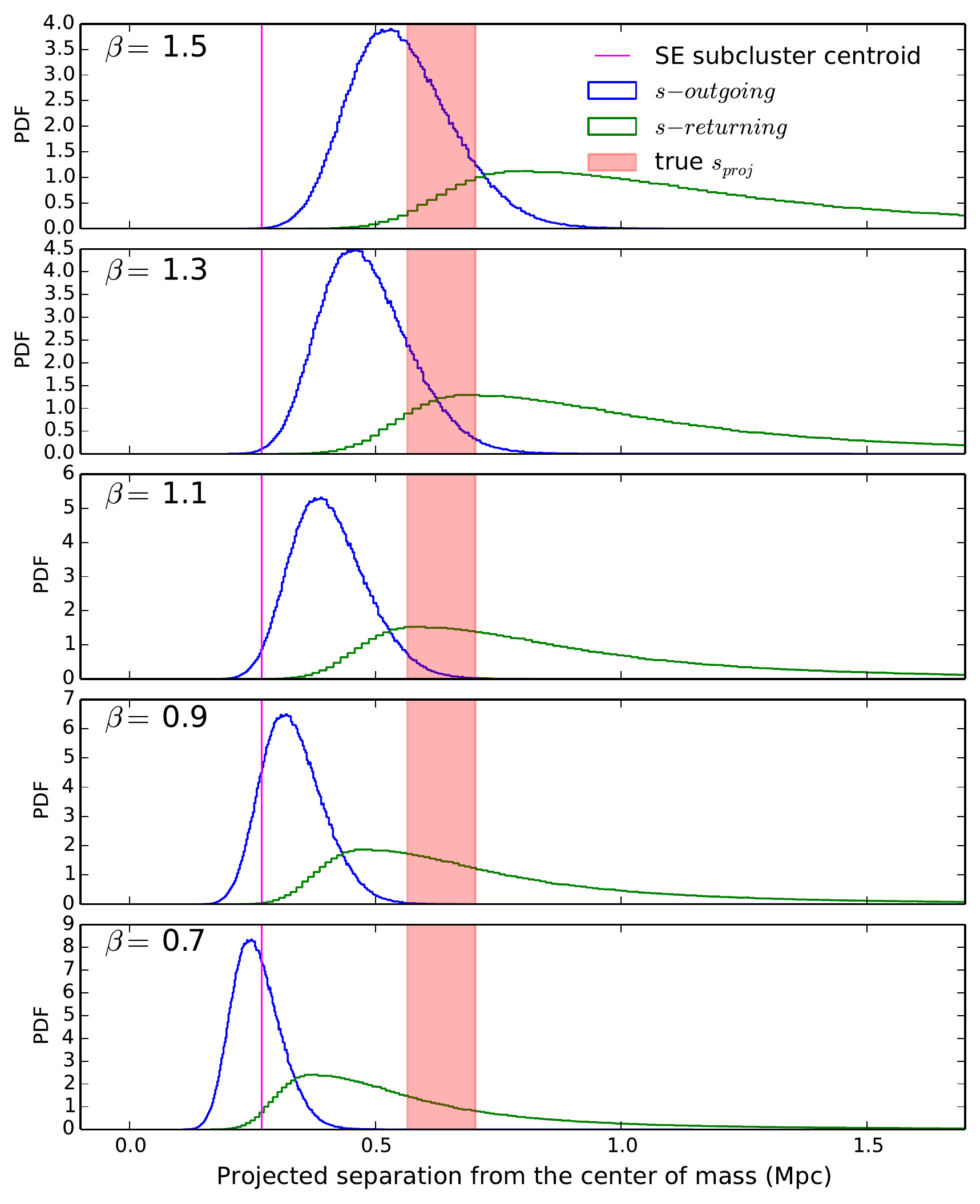}
	\caption{Comparison of the PDFs of the observed position of the NW relic (red bar
		includes 95\% confidence interval of location of relic in the center of
		mass frame) with the	predicted position from the two simulated merger scenarios (blue for outgoing and green for the returning scenario). 
	For the plausible values of $\beta  < 1.1$, the returning model is
	preferred. For comparison purpose, we also show that $\beta > 1.3$ (top
	panel) for the outgoing scenario to be favored.  
	Note that we made use of the polarization weight for producing this figure. 
	\label{fig:positionprior}}
\end{figure}
\clearpage
\bsp 
\label{lastpage} 
\end{document}